\documentclass[%
 reprint,
superscriptaddress,
longbibliography,
 amsmath,amssymb,
 aps,
]{revtex4-2}

\usepackage{graphicx}
\usepackage{dcolumn}
\usepackage{bm}
\usepackage{float}
\usepackage[normalem]{ulem}
\usepackage{microtype}
\usepackage[dvipsnames]{xcolor}
\usepackage[breaklinks]{hyperref}

\usepackage{siunitx}
\DeclareSIUnit\sq{sq}
\DeclareSIUnit\T{T}
\DeclareSIUnit\dBm{dBm}

\begin{document}


\title{From nonreciprocal to charge-4e supercurrents in Ge-based Josephson devices with tunable harmonic content}

\author{Axel Leblanc}
\email{E-mail: axel.leblanc@cea.fr}
\affiliation{Univ. Grenoble Alpes, CEA, Grenoble INP, IRIG, PHELIQS, 38000 Grenoble, France}
\author{Chotivut Tangchingchai}
\affiliation{Univ. Grenoble Alpes, CEA, Grenoble INP, IRIG, PHELIQS, 38000 Grenoble, France}
\author{Zahra Sadre Momtaz}
\affiliation{Institut Néel, CNRS/UGA, Grenoble 38042, France}
\author{Elyjah Kiyooka}
\affiliation{Univ. Grenoble Alpes, CEA, Grenoble INP, IRIG, PHELIQS, 38000 Grenoble, France}
\author{Jean-Michel Hartmann}
\affiliation{Univ. Grenoble Alpes, CEA, LETI, 38000 Grenoble, France}
\author{Gonzalo Troncoso Fernandez-Bada}
\affiliation{Univ. Grenoble Alpes, CEA, Grenoble INP, IRIG, PHELIQS, 38000 Grenoble, France}
\author{Boris Brun-Barriere}
\affiliation{Univ. Grenoble Alpes, CEA, Grenoble INP, IRIG, PHELIQS, 38000 Grenoble, France}
\author{Vivien Schmitt}
\affiliation{Univ. Grenoble Alpes, CEA, Grenoble INP, IRIG, PHELIQS, 38000 Grenoble, France}
\author{Simon Zihlmann}
\affiliation{Univ. Grenoble Alpes, CEA, Grenoble INP, IRIG, PHELIQS, 38000 Grenoble, France}
\author{Romain Maurand}
\affiliation{Univ. Grenoble Alpes, CEA, Grenoble INP, IRIG, PHELIQS, 38000 Grenoble, France}
\author{\'Etienne Dumur}
\affiliation{Univ. Grenoble Alpes, CEA, Grenoble INP, IRIG, PHELIQS, 38000 Grenoble, France}
\author{Silvano De Franceschi}
\affiliation{Univ. Grenoble Alpes, CEA, Grenoble INP, IRIG, PHELIQS, 38000 Grenoble, France}
\author{François Lefloch}
\email{E-mail: francois.lefloch@cea.fr}
\affiliation{Univ. Grenoble Alpes, CEA, Grenoble INP, IRIG, PHELIQS, 38000 Grenoble, France}

\date{\today}

\begin{abstract}
Hybrid superconductor(S)-semiconductor(Sm) devices bring a range of new functionalities into superconducting circuits. In particular, hybrid parity-protected qubits and Josephson diodes were recently proposed and experimentally demonstrated. Such devices leverage the non-sinusoidal character of the Josephson current-phase relation (CPR) in highly transparent S-Sm-S junctions.  Here we report an experimental study of superconducting  quantum-interference devices (SQUIDs) embedding Josephson field-effect transistors fabricated from a SiGe/Ge/SiGe heterostructure grown on a 200-mm silicon wafer. The single-junction CPR shows up to three harmonics with gate tunable amplitude. In the presence of microwave irradiation, the ratio of the first two dominant harmonics, corresponding to single and double Cooper-pair transport processes, is consistently reflected in relative weight of integer and half-integer Shapiro steps. A combination of magnetic-flux and gate-voltage control enables tuning the SQUID functionality from a nonreciprocal Josephson-diode regime with 27 \% asymmetry to a $\pi$-periodic Josephson regime suitable for the implementation of parity-protected superconducting qubits. These results illustrate the potential of Ge-based hybrid devices as versatile and scalable building blocks of novel superconducting quantum circuits.    

\end{abstract}

\maketitle

\section{Introduction}

The recent years have seen a revival of interest toward superconductor(S)-semiconductor(Sm) devices. These hybrid systems leverage, on the one hand,  the macroscopic quantum coherence coming from superconductivity and, on the other hand, the field-effect charge control enabled by semiconductor materials. A typical example is the Josephson field-effect transistor (JoFET), a three-terminal device consisting of two superconducting contacts connected by a gate-tunable semiconducting channel. Owing to the superconducting proximity effect, a dissipationless supercurrent can flow through the normal-type semiconductor channel,  with a maximal value, the so-called critical current, that depends on the applied gate voltage.  A variety of JoFETs have been realized using different semiconductor materials. Some of these JoFETs were used to realize gate-tunable superconducting qubits often called gatemons \cite{larsen_semiconductor-nanowire-based_2015, de_lange_realization_2015, casparis_gatemon_2016, casparis_superconducting_2018, wang_coherent_2019} and parametric amplifiers \cite{butseraen_gate-tunable_2022, sarkar_quantum_2022, phan_gate-tunable_2023, splitthoff_gate-tunable_2023, strickland_characterizing_2023, huo_gatemon_2023, hertel_gate-tunable_2022, zhuo_hole-type_2023}. 

In a JoFET, high-transparency S-Sm contacts enable the phase coherent transfer of $m$ Cooper pairs at a time, resulting in the emergence of $\sin(m\varphi)$ components in the current phase relation (CPR) where $\varphi$ is the phase difference between the two superconductors \cite{haberkorn_theoretical_1978}. The gate permits to fine tune the critical current amplitude as well as the CPR harmonic composition \cite{spanton_currentphase_2017}. 

In the context of superconducting qubits, the ability to tailor the CPR provides a means to engineer the support of the wavefunctions encoding the qubit.  This idea has lead to  parity-protected qubits leveraging  supercurrents carried by correlated pairs of Cooper pairs in $\sin(2\varphi)$ Josephson elements \cite{gladchenko_superconducting_2009, doucot_physical_2012,smith_superconducting_2020, smith_magnifying_2022}. Along the same line, $\sin(2\varphi)$ Josephson elements for parity protection were recently obtained using S-Sm-S junctions with either an InAs  \cite{larsen_parity-protected_2020, ciaccia_charge-4e_2023} or, as in this work, a Ge channel  \cite{valentini_parity-conserving_2023} 

Here we investigate the CPR of a Ge-based Josephson junction using an asymmetric SQUID featuring a wide and a narrow JoFET~\cite{della_rocca_measurement_2007}. The CPR Fourier transform exhibits gate-tunable higher harmonics, revealing charge-2e, charge-4e and charge-6e dissipationless transport. Notably, we observe the possibility of suppressing the higher harmonics by polarizing the JoFET near pinch-off. Subsequently, by investigating the response to microwave excitation,  we observe half-integer Shapiro steps corroborating the existence of charge-4e supercurrents. 

It was recently proposed \cite{souto_josephson_2022} that non-sinusoidal CPR in nearly symmetric SQUIDs can lead to asymmetric transport characteristics. This behavior, commonly known as superconducting diode effect (SDE), requires time-reversal symmetry broken by a magnetic flux and is highly tunable unlike previously reported realizations based on inversion symmetry breaking \cite{ando_observation_2020,bauriedl_supercurrent_2022,daido_intrinsic_2022,yun_magnetic_2023,ilic_theory_2022,he_phenomenological_2022, matsuo_josephson_2023}, spin orbit interaction coupled to a Zeeman field \cite{yokoyama_anomalous_2014,zazunov_anomalous_2009,halterman_supercurrent_2022,strambini_superconducting_2022} or magnetic JJ \cite{pal_quantized_2019}. The diode efficiency vanishes when the SQUID operates in a perfectly symmetric regime, i.e. when the two junctions have the same CPR, or when the SQUID is flux-biased at half the flux quantum. By harnessing the SDE within a SQUID made of two similar JoFETs (we observe diode efficiency up to 27\%), we identify its regime of critical current balance. In this configuration, Shapiro steps measurements at half flux quantum bias evidences a pronounced reduction in the odd harmonics due to destructive interference within the SQUID. 

This realization yields a device primarily governed by a charge-4e supercurrent, effectively creating a $\sin(2\varphi)$ Josephson element, thus opening the pathway for the development of Ge based parity-protected qubits.

\section{Device and material properties}

The initial heterostructure consists of a $\mathrm{Si}_{0.21}\mathrm{Ge}_{0.79}$ virtual substrate (with a 10\%Ge/µm grading from a few \% up to 79\% of Ge and a 2.5 µm thick Si0.21Ge0.79 layer on top) grown on a Si(001) wafer. A \SI{16}{\nm} thick Ge quantum well is grown on top of the polished virtual substrate and covered by a \SI{22}{\nm} thick $\mathrm{Si}_{0.21}\mathrm{Ge}_{0.79}$ layer and a \SI{2}{\nm} thick Si cap layer.  In accumulation regime the hole mobility $\mu_h$ measured in $\qtyproduct[product-units = power]{60 x 600}{\um}$ Hall bars, reaches $\SI{1e5}{\square\cm\per\V\per\s}$ while the elastic mean free path is estimated to $l_e \approx \SI{1.5}{\um}$ \cite{hartmann_epitaxy_2023}.

The fabrication of the Ge JoFETs starts with the etching of a long mesa (typically $\SI{5}{\um}$) with various widths ranging from $\SI{1}{\um}$ to $\SI{10}{\um}$. In a second step, the superconducting contacts are defined by locally etching the Si cap layer and the SiGe top layer, followed by the ex-situ deposition and lift-off of a \SI{50}{\nm} thick Al layer in contact with the Ge quantum well. The aluminium contact has a typical critical temperature of \SI{1.54}{\K}. An AlOx oxide is then deposited everywhere using Atomic Layer Deposition (ALD). To finish the process, a Ti/Au top gate covers the Ge central channel and slightly overlaps the Al contacts (Fig1. a). 

The asymmetric SQUID consists of two Al-Ge-Al JoFETs placed in an Al loop (Fig. 1b).  The narrow junction $\mathrm{J_n}$ has a width of \SI{1}{\um}, while the wide junction $\mathrm{J_w}$ is \SI{8}{\um} wide. Both junctions are \SI{300}{\nm} long.  In the following experiments, the devices are cooled down to a base temperature of \SI{35}{\milli\K}.
To measure the electrical behavior of $\mathrm{J_n}$ (resp. $\mathrm{J_w}$) independently, we apply a large gate voltage on $\mathrm{J_w}$ (resp. $\mathrm{J_w}$) (typically $\SI{+4}{\V}$) to suppress the conductance on the respective arm of the SQUID. Both junctions can be fully proximitized and show a non-dissipative current up to a certain switching current (identified as the critical current) that is gate tunable. At full accumulation ($V_G = \SI{-2}{V}$), the critical currents are $I_C^\mathrm n = \SI{158}{\nA}$ and $I_C^\mathrm w = \SI{2.39}{\uA}$ with very remarkable similar gate dependency (Fig. 1c). The $IcR_N$ product of both junctions are of the order of $\SI{100}{\micro\eV}$ and very close to the state-of-the art \cite{hendrickx_ballistic_2019, vigneau_germanium_2019, aggarwal_enhancement_2021, tosato_hard_2023, xiang_gesi_2006}.  

The symmetric SQUID device used to study the SDE is composed of two nearly identical junctions with W=\SI{4}{\um} and L=\SI{300}{\nm} and shows similar tunable critical current at low temperature.  

\begin{figure}[h]
\includegraphics[width=0.48\textwidth]{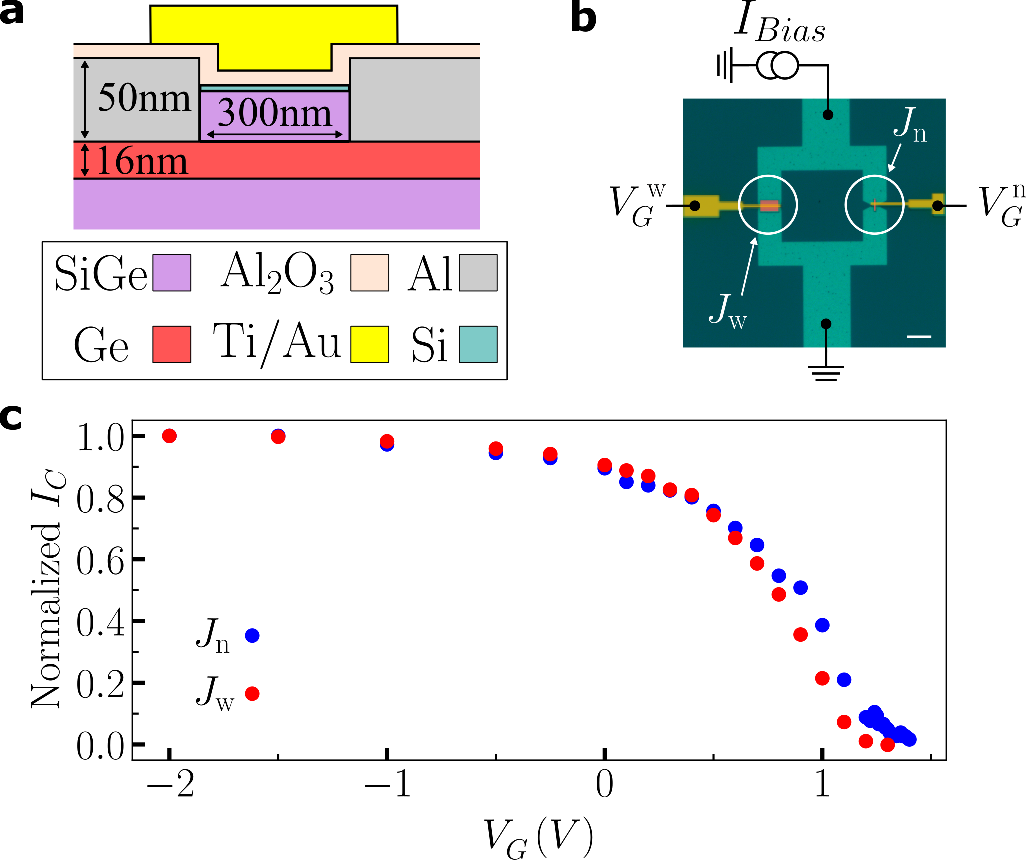}
\caption{\textbf{Ge/Al Josephson field effect transistors (JoFET) placed in a SQUID.}  \textbf{a}, Cross-section sketch of a JoFET where the Ge quantum well is contacted by two superconducting aluminium leads. The Ti/Au top gate allows to modify the hole carrier density in the channel. \textbf{b}, False color SEM image of the SQUID with a wide (left) and a narrow (right) junctions placed in an aluminium loop (green). Electrostatic gates are shown in yellow. The scale bar is $\SI{10}{\um}$. \textbf{c}, Critical current of the two junctions normalised to their values at $V_G = \SI{-2}{V}$.}
\label{fig:characterization}
\end{figure}

\section{Current-phase relation of a single G\lowercase{e}/A\lowercase{l} JoFET}

\begin{figure*}[]
\includegraphics[width=1\textwidth]{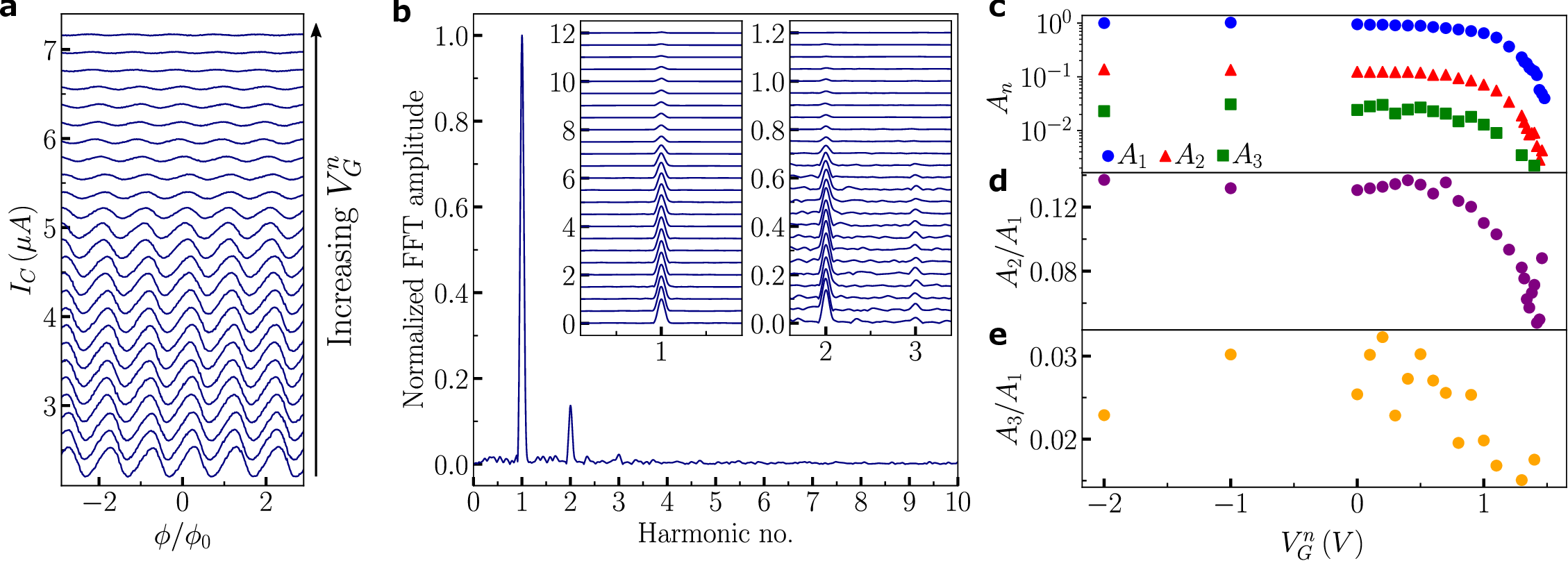}
\caption{\textbf{Gate modulation of the CPR harmonic content.}  \textbf{a}, $I_c$ modulations (shifted for clarity) of the asymmetric SQUID as a function of the flux for various $V_G^\mathrm n$ (each line correspond to one point in \textbf{c}, \textbf{d} , \textbf{e}) from full accumulation ($V_G^\mathrm n=\SI{-2}{\V}$) to near threshold ($V_G^\mathrm n=\SI{1.48}{\V}$) with a fixed gate voltage applied to the wide junction ($V_G^\mathrm w=\SI{-2}{\V}$).  \textbf{b}, Fast Fourier transform (FFT) (normalized to the first harmonic amplitude) of $I_C(\phi)$  at $V_G^\mathrm n=\SI{-2}{\V}$ and, in the insets, for the same $V_G^\mathrm n$ range as in \textbf{a} (shifted for clarity). The three first harmonics are clearly visible and higher harmonics disappear faster when $V_G^\mathrm n$ gets close to the threshold voltage. \textbf{c}, Amplitude of the first three harmonics $A_1$, $A_2$ and $A_3$ (in log scale) versus the gate voltage $V_G^\mathrm n$. All of them tends to decrease with gate voltage since the overall flux modulation of $I_C$ decreases. \textbf{d[e]}, Ratio between the second [third] and the first harmonic amplitudes. The ratios decreasing highlights the disappearance of higher harmonics before the first one and thus a transition from a multi harmonics CPR toward a more sinusoidal one.}
\label{fig:CPR}
\end{figure*}

The loop dimensions of the asymmetric SQUID used for CPR measurements have been chosen large enough to avoid Fraunhofer reduction in the flux range of interest (see \ref{sup_stat}). The width of the loop arms is \SI{10}{\um} and the associated self-inductance is estimated to be \SI{102}{\pico\henry}. Its contribution to the measurement is negligible and discussed in \ref{sup_ind}.

The total supercurrent $I_\mathrm{SQUID}$ flowing through the SQUID is the sum of the current flowing through each junction and thus includes the two current phase relations $I_\mathrm w(\varphi_\mathrm w)$ and $I_\mathrm n(\varphi_\mathrm n)$.  Since one junction is much wider than the other, its critical current $I_C^\mathrm w$ is also much larger than that of the narrow junction $I_C^\mathrm n$ ($I_C^\mathrm w/I_C^\mathrm n = 15$). Thus, the maximum supercurrent through the SQUID is reached at $\varphi_\mathrm w \approx \pi/2$ i.e. when the maximum current through $\mathrm{J_w}$ is reached. Taken into account the fluxoid relationship that relates the superconducting phases to the applied magnetic flux, the total critical current flux dependency of the  SQUID can be written as :

\begin{equation}
I_C^{SQUID} = I_\mathrm w(\pi/2) + I_\mathrm n(\pi/2 + 2\pi \Phi_{ext}/\Phi_0)
\label{SQUID_eq}
\end{equation}

where $\Phi_\mathrm{ext}$ is the magnetic flux threading the loop and $\Phi_0=h/2e$ the superconducting flux quantum.

One can clearly see that measuring the critical current of the whole SQUID $I_C^\mathrm{SQUID}$ versus the applied flux $\Phi_\mathrm{ext}$ allows to probe the current phase relation of the narrow junction $\mathrm{J_n}$ \cite{della_rocca_measurement_2007} (see \ref{sup_stat} for details on the measurement protocol). 

For this experiment, the wide junction $\mathrm{J_w}$ is kept in the full accumulation regime ($V_G^\mathrm w=\SI{-2}{\V}$).  As expected, oscillations of the SQUID critical current are observed, as well as a background curvature associated to the Fraunhofer effect on the wide junction (\ref{sup_stat}). The flux modulation $I_{C}(\Phi_\mathrm{ext})$, studied for different values of the gate voltage $V_G^\mathrm n$ and shown in Fig. 2a after numerically removing the background, is a direct measurement of the CPR of the narrow junction.

One can already notice that $I_{C}(\Phi_\mathrm{ext})$ is skewed in the highly accumulated regime suggesting the presence of higher harmonics.  
In order to better evidence the skewness and to quantify the various components of the CPR, we applied a Fourier transform to these data over 25 periods (Fig. 2b).

The spectral decomposition at full accumulation ($V_G^\mathrm n=\SI{-2}{\V}$) is shown in Fig. 2b. The result shows a clear contribution from the first and second harmonics, and also from the third harmonic but with less pronounced evidence. This confirms that the CPR is not purely sinusoidal. The insets of Fig. 2b show the gate dependence of the three harmonics. The first harmonic survives up to the largest gate voltages applied ($V_G^\mathrm n = \SI{1.48}{\V}$) while the second and third harmonics vanish at a lower gate voltage.

The amplitude of the three first harmonics are extracted by taking the height of each peak in the Fourier decomposition. $A_n$ is the weight of the $n^{th}$ harmonic and is plotted as a function of $V_G^n$ in Fig. 2c. When $V_G^\mathrm n$ approaches the threshold voltage ($V_{th}\approx \SI{1.5}{\V}$), we observe a reduction of all harmonic amplitudes consistent with the reduction of the critical current. The ratio between $A_2$ (resp. $A_3$) and $A_1$ shown in Fig. 2d (resp. Fig. 2e) goes up to 0.13 (resp. 0.03). Interestingly, the gate dependence reveals the faster disappearance of higher harmonics compare to the main one. 
As mentioned above, higher harmonics in the CPR are intrinsic to the proximitized Josephson junctions. We have checked (see \ref{sup_squid}) that multiple harmonics cannot originate from arm inductances as it is the case with S-I-S junctions. 
The faster disappearance of the higher harmonics reveals a change in the CPR as the proximitized Ge region gets depleted from carriers.  

This is consistent with the decrease of the number of conduction channels (as the Fermi wavelength increases) leading to the reduction of the total non-dissipative current. Together, the Fermi velocity decreases and so is the proximity effect (through the reduction of penetration of Andreev pairs in the Ge channel). Moreover, the transmission coefficients of the conducting modes and the transparency of the S-Sm interfaces are affected by the reduction of carriers. All these effects contribute to decrease the probability of (multiple) transfer of Cooper pairs \cite{haberkorn_theoretical_1978,golubov_current-phase_2004}.

\section{Shapiro steps in a single G\lowercase{e}/A\lowercase{l} JoFET}

In the previous section, we have shown the results of the direct measurement of the CPR of the narrow junction. Thanks to the high gate control of our Ge based JoFET, the wide junction of the exact same SQUID device can be tuned in the very depleted regime at $V_G^\mathrm w=\SI{4}{\V}$ where the conductance is zero. The device is now only composed of the narrow junction for which the CPR is known.

In the following, we focus on the AC Josephson effect of the narrow junction, which should highlight the presence of higher harmonics in the CPR. Under microwave irradiation at frequency $f$, the current-voltage characteristic shows steps at some specific voltages $V=nhf/2e$ where $h$ is the Plank constant, $e$ the elementary charge and n = 0,1,2,.. called Shapiro steps \cite{shapiro_josephson_1963}. If the CPR of the Josephson junction contains a second harmonic, half-integer Shapiro steps also appear (n=1/2,3/2,..) indicating the coherent transport of pairs of Cooper pairs through the junction (i.e. a charge-4e supercurrent)\cite{ueda_evidence_2020,iorio_half-integer_2023}.

Fig. 3a shows the differential resistance measurement (red line) and the measured DC voltage (blue line) as a function of the current bias under a radio-frequency irradiation at $f=\SI{3.05}{\giga\Hz}$ and with a fixed microwave power $P_{RF}=\SI{-12}{\dBm}$. We identify clear dips in the differential resistance that correspond to integer and half-integer Shapiro steps (highlighted by dashed grey lines). 

Fig. 3b shows the differential resistance as a function of the microwave drive power $P_{RF}$ and the DC bias $I_{DC}$. Dark regions correspond to dips in the differential resistance and are associated with Shapiro steps. The widths of these steps follow the usual Bessel-like behavior \cite{russer_influence_1972}. The clear observation of half-integer steps is perfectly consistent with previously measured harmonic content of the CPR.  

By applying more positive gate voltages ($V_G^\mathrm n$ near threshold), one would expect half integer steps to disappear before integer ones as the CPR becomes purely sinusoidal. The quantitative analysis of this effect is challenging because of the overlap of the different step signals for such rounded steps. Shapiro patterns for different gate voltages are shown in \ref{sup_shap}.

As an intermediate conclusion, we have shown that proximitized planar SiGe based Josephson junctions exhibit intrinsic non purely sinusoidal current phase relation. This response has been probed and measured in a single junction, first in a SQUID geometry to directly measure the CPR and under radio-frequency irradiation to reveal half-integer Shapiro steps. This has been made possible thanks to the high gate tunability of the SiGe JoFETs.   

\begin{figure}[]
\includegraphics[width=0.5\textwidth]{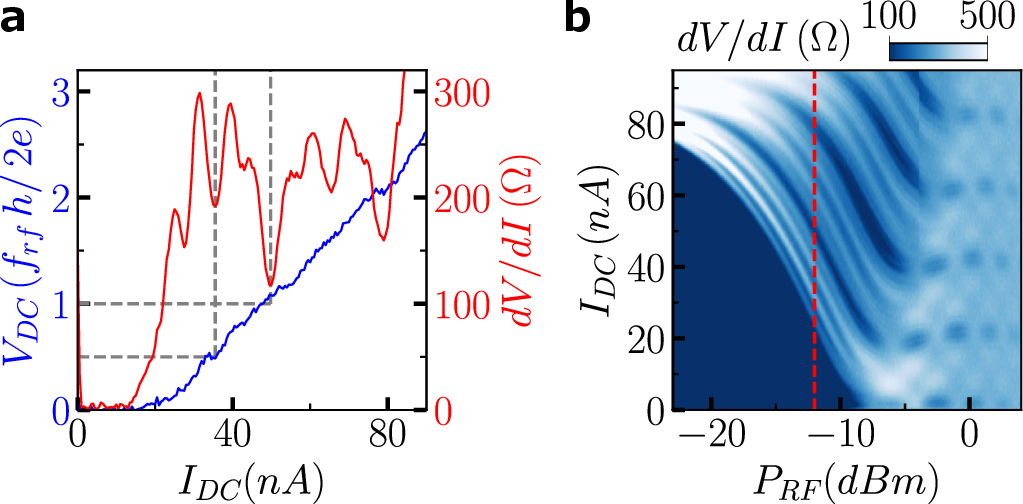}
\caption{\textbf{Integer and half-integer Shapiro steps.}  \textbf{a}, DC voltage normalized to the expected Shapiro steps positions (blue line) and differential resistance (red line) as function of the bias current $I_{DC}$ under radio-frequency irradiation at $f=\SI{3.05}{\giga\Hz}$ with a power $P_{RF}=\SI{-12}{\dBm}$ (see the dashed red line in \textbf{b}). The dashed grey lines are guides to the eye for the first half integer and integer steps. \textbf{b}, Differential resistance as function of the bias current $I_{DC}$ as well as the microwave power $P_{RF}$. Integer (wide dark lines) and half-integer (narrow dark lines) follow the Bessel like oscillations as a function of the RF power.}
\end{figure}

\section{Superconducting diode effect in a symmetric SQUID}

In a SQUID with negligible arms inductances, superconducting diode effect (SDE) can still emerge if at least one of the junctions has a non-sinusoidal CPR \cite{souto_josephson_2022,ciaccia_gate-tunable_2023-1,valentini_parity-conserving_2023}. To investigate this possibility, we have fabricated a SQUID with two nominally identical JoFETs,  $\mathrm{J_1}$ and $\mathrm{J_2}$ with W=\SI{4}{\um} and L=\SI{300}{\nm}. The diode effect arises if (i) the external flux is not an integer multiple of the half flux quantum $\Phi = n\Phi_0/2$ and (ii) the two CPRs of the two junctions are not perfectly balanced. The SDE can be quantified through the diode efficiency:

\begin{equation}
\eta=\frac{I_C^+ - I_C^-}{I_C^+ + I_C^-}
\label{diode_eq}
\end{equation}

where $I_C^+$ and $I_C^-$ are the forward and backward critical currents of the  SQUID. 

Fig. 4a shows the diode efficiency $\eta$ as a function of the gate voltage of $J_1$ ($V_G^1$) and the normalized flux $\Phi_{ext}/\Phi_0$ for a fixed value of $V_G^2=\SI{1.75}{\V}$ together with two $I_c^{+(-)}$  traces (Fig. 4c,d) from which the efficiency is calculated. As expected, the diode efficiency vanishes at half the flux quantum and follows very well the predicted behavior \cite{souto_josephson_2022} with a maximum that goes up to 27\%. At $V_G^1=\SI{1.825}{\V}$ (Fig. 4c) the diode efficiency drops to zero for any flux value. This suggests that for this combination of gate voltages the SQUID is perfectly symmetric. But this gate values combination is not unique. 

To find every gate configurations that perfectly balance the SQUID, the device is flux-biased slightly below the half flux quantum ($\Phi_\mathrm{ext} = 0.45 \Phi_0$) and the diode efficiency measured as a function of $V_G^\mathrm 1$ and $V_G^\mathrm 2$ . The result (Fig. 4b) highlights the existence of a line (in white) corresponding to all gate voltage combinations where the SQUID operates in a perfectly balanced regime. The critical current of the SQUID along this line is shown in \ref{sup_sweet_line}. The fact that this line does not follow the $V_G^\mathrm 1=V_G^\mathrm 2$ diagonal means that the two JoFETs have fairly close yet not identical characteristics. Nevertheless, the symmetric situation can be easily obtained thanks to the independent tunability of the junctions.

\begin{figure}[]
\includegraphics[width=0.48\textwidth]{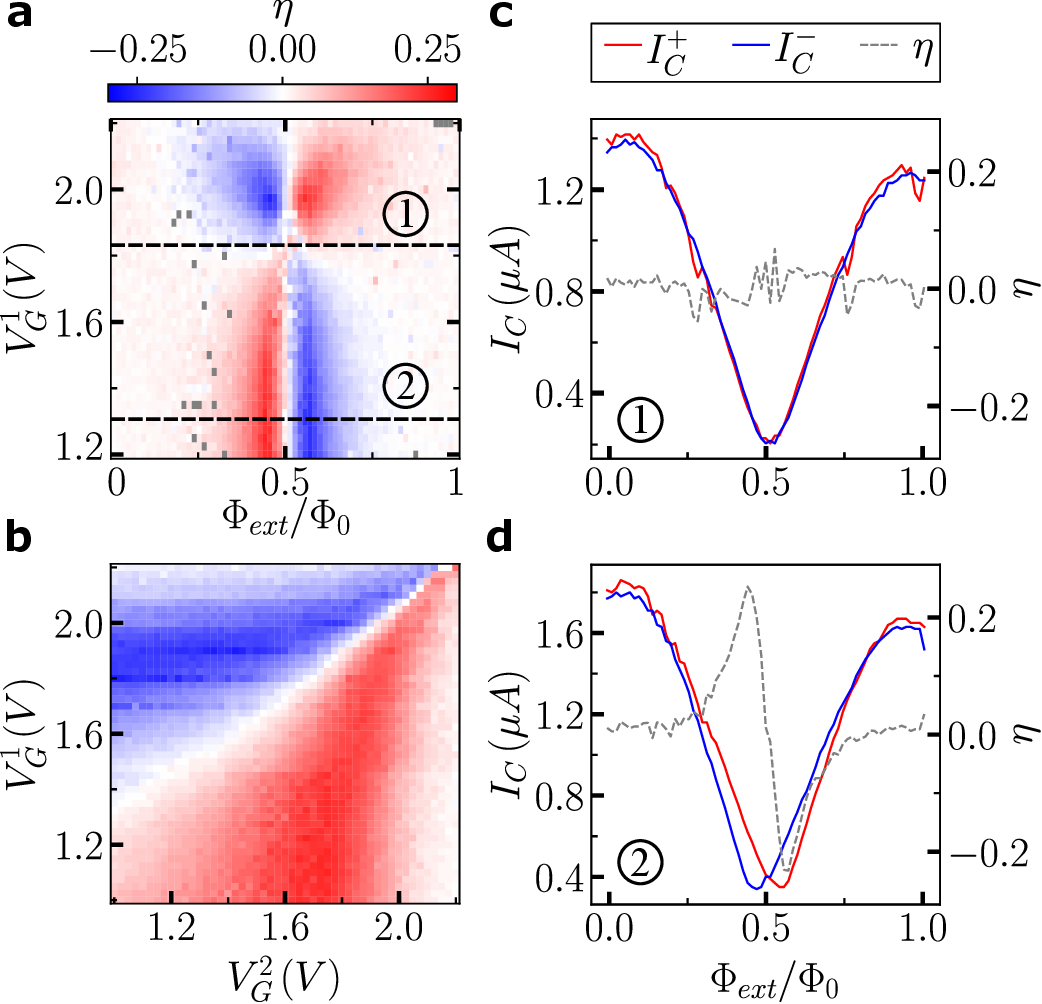}
\caption{\textbf{Gate and flux dependence of the superconducting diode effect.} \textbf{a}, Diode efficiency $\eta$ as function of the normalized applied flux $\Phi_\mathrm{ext}/\Phi_0$ and $V_G^1$ for a fixed $V_G^2=\SI{1.75}{\V}$. $\eta$ vanishes at half flux quantum $\Phi_\mathrm{ext}/\Phi_0 = 0.5$ and for the peculiar gate voltage $V_G^1 = \SI{1.825}{\V}$. \textbf{b}, Diode efficiency as function of the two gate voltages $V_G^1$ and $V_G^2$ for a fixed flux biasing slightly below the half flux quantum $\Phi_\mathrm{ext}/\Phi_0 = 0.45$. The white line corresponds to perfectly symmetric regime where the SDE disappears. \textbf{c,d} Forward $I_c^+$ (red line) and backward $I_c^-$ (blue line) critical currents and extracted diode efficiency $\eta$ as a function of the normalized applied flux $\Phi_{ext}/\Phi_0$ for $V_G^1=\SI{1.75}{\V}$ and $V_G^1=\SI{1.3}{\V}$.}
\end{figure}

\section{Shapiro steps in a perfectly symmetric SQUID }

In the previous section we have shown that the superconducting diode effect (SDE) observed in a symmetric SQUID is consistent with the non sinusoidal CPR of the junctions. We have also demonstrated that the contribution of the various harmonics strongly depends on the applied magnetic flux threading the SQUID. Moreover, as detailed before, DC Shapiro steps observed under radio-frequency irradiation can also reveal CPR multiple harmonics. 
In this section, we present the results of Shapiro steps measured in the exact same SQUID than the one used to reveal SDE. To do so, we operate the SQUID for a fixed set of gate voltages $V_G^\mathrm 1=\SI{1.825}{\V}$ and $V_G^\mathrm 2=\SI{1.75}{\V}$,  which correspond to one of the symmetric regime combinations and vary the magnetic flux. 

Fig. 5a is the theoretical amplitudes of the first and second harmonics as a function of the flux threading in a symmetric SQUID (see \ref{sup_simu}). Fig. 5b shows the differential resistance of the SQUID as a function of the applied external flux $\Phi_\mathrm{ext}$ and the DC voltage $V_{DC}$ normalized to the expected Shapiro steps positions, for a given microwave radiation power $P_{RF}=\SI{-8}{\dBm}$ and frequency $f = \SI{7}{\giga\Hz}$. Dark blue ridges reveal the position of the Shapiro steps and evidence the emergence of half integer steps around $\Phi_0/2$. Fig. 5c,d,e show the differential resistance as a function of the microwave radiation power and the normalized DC voltage for three different fluxes $\Phi_\mathrm{ext}=0$, $\Phi_\mathrm{ext}=\Phi_0/2$ and $\Phi_\mathrm{ext}=\Phi_0/4$ corresponding to the dashed grey lines in Fig. 5a. At $\Phi_\mathrm{ext}=\Phi_0$, both integer and half-integer steps are visible with a stronger intensity of the integer steps and follows the Bessel-type oscillations as a function of the power. At $\Phi_\mathrm{ext}=\Phi_0/2$ integer and half-integer steps are still visible but with a relative contrast that is weaker than for $\Phi_\mathrm{ext}=0$. This reduction is consistent with the vanishing of the $1^\mathrm{st}$ harmonic at $\Phi_\mathrm{ext}=\Phi_0/2$ but where integer steps correspond to a multiple photons process of half-integer steps. At $\Phi_\mathrm{ext}=\Phi_0/4$ however, the contribution of the $2^\mathrm{nd}$ CPR harmonic vanishes and the half-integer steps disappear completely. Therefore, we demonstrated the flux-tunability of the $1^\mathrm{st}$ and $2^\mathrm{nd}$ harmonics opening the possibility to engineer a $\sin(2\varphi)$ Josephson element dominated by charge-4e supercurrent.

\begin{figure}[]
\includegraphics[width=0.48\textwidth]{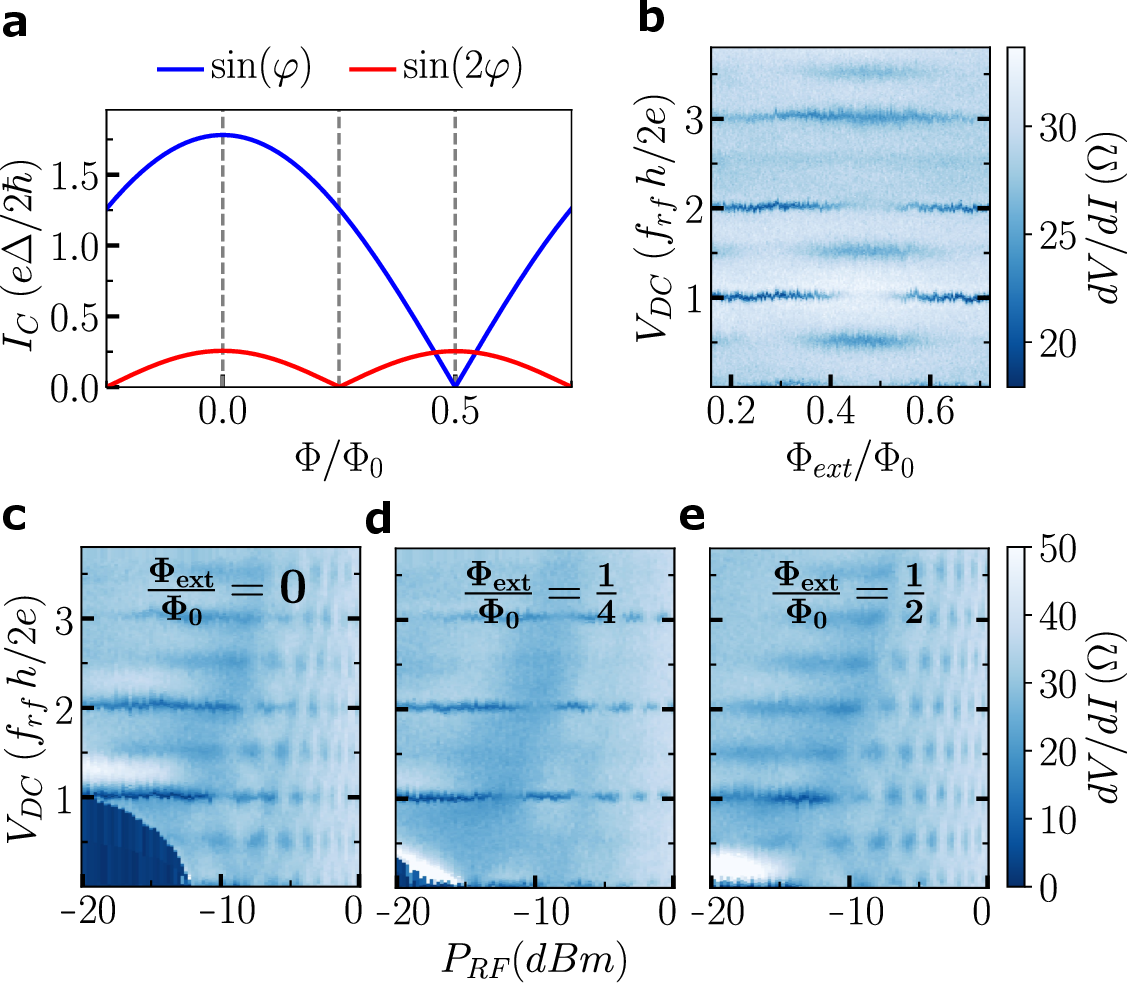}
\caption{\textbf{Half integer Shapiro steps as a probe for $\mathbf{sin(2\varphi)}$ CPR.} \textbf{a}, Computation of the two first harmonics amplitudes of a symmetric SQUID CPR for a single channel short junction model with a transparency $\tau=0.7$ (see \ref{sup_simu}). \textbf{b}, Differential resistance as a function of the DC voltage $V_{DC}$ normalized to the Shapiro steps expected positions and the applied flux normalized to $\phi_0$ for a fixed $P_{RF}=\SI{-8}{dBm}$ and in the balanced regime ($V_G^1=\SI{1.825}{\V}$ and $V_G^2=\SI{1.75}{\V}$). Dark lines correspond to differential resistance dips associated to Shapiro steps. \textbf{b}[\textbf{d}][\textbf{e}], Differential resistance as a function of the normalized DC voltage and the RF excitation power at fixed applied flux $\Phi_\mathrm{ext}=0$ [$\Phi_\mathrm{ext}=\Phi_0/2$][$\Phi_\mathrm{ext}=\Phi_0/4$].}
\end{figure}

\section{Conclusion}
In conclusion, our work harnesses the high gate tunability offered by SiGe-based proximitized Josephson junctions to conduct comprehensive measurements of the CPR and Shapiro steps within a single JoFET. The CPR clearly reveals gate-tunable non-purely sinusoidal phase dependence, reveling higher-order harmonics that correspond to the coherent transfer of both single and multiple Cooper pairs. This observation is supported by the presence of both integer and half integer Shapiro steps. We also revealed a gate and flux-tunable SDE in a SQUID geometry, made possible once again by the JoFETs high transparency and gate tunability \cite{souto_josephson_2022}. Ultimately, we demonstrated the realization of a so-called $\sin(2\varphi)$ Josephson element in a SQUID biased at half flux quantum and finely gate-tuned to its balanced critical current point. Our results underscore the remarkable potential of the SiGe-based heterostructures for the realization of parity-protected superconducting qubits without the need for large inductance engineering as required with tunnel S-I-S junctions\cite{larsen_parity-protected_2020,smith_superconducting_2020,schrade_protected_2022}. Our study focused on wide junctions, ranging from 1 to \SI{10}{\um}, with critical current spanning from a few hundred \SI{}{\nA} to a few \SI{}{\uA}. These values suggest that narrower junctions, probably necessary to avoid Andreev bound states at low energies, will still have a large enough Josephson energy for the realization of suitable Gatemon qubits. Moreover, such materials are fully compatible with the CMOS technology and potentially  large scale integration.

\begin{acknowledgments}
This work has been supported by the ANR project SUNISIDEuP (ANR-19-CE47-0010), the PEPR ROBUSTSUPERQ (ANR-22-PETQ-0003), the ERC starting grant LONGSPIN (Horizon 2020 - 759388) and the Grenoble LaBEX LANEF. We thank Frederic Gustavo and Jean-Luc Thomassin for their contribution in the nanofabrication at the PTA (CEA-Grenoble) and the CNRS Neel Institute for the access to the Nanofab facility.
\end{acknowledgments}

\bibliography{biblio} 

\widetext
\clearpage
\begin{center}
\textbf{\large Supplemental Materials}
\end{center}
\setcounter{equation}{0}
\setcounter{figure}{0}
\setcounter{table}{0}
\setcounter{section}{0}
\setcounter{page}{1}
\makeatletter
\renewcommand{\theequation}{S\arabic{equation}}
\renewcommand{\thefigure}{S\arabic{figure}}
\renewcommand{\thetable}{S\arabic{table}}
\renewcommand{\thesection}{S-\Roman{section}}

\section{Asymmetric SQUID critical current and normal resistance} \label{sup_IcRn}

The critical current is defined as the switching current from the superconducting to the normal state. The data presented in Fig. S1 have been measured at a not optimized perpendicular magnetic field, and explain the $I_C$ reduction compared to those in the main text. The normal resistances are measured at large bias ($eV_{Bias} > 3\Delta_{Al}$). 

\begin{figure}[H]
\centering
\includegraphics[width=0.8\textwidth]{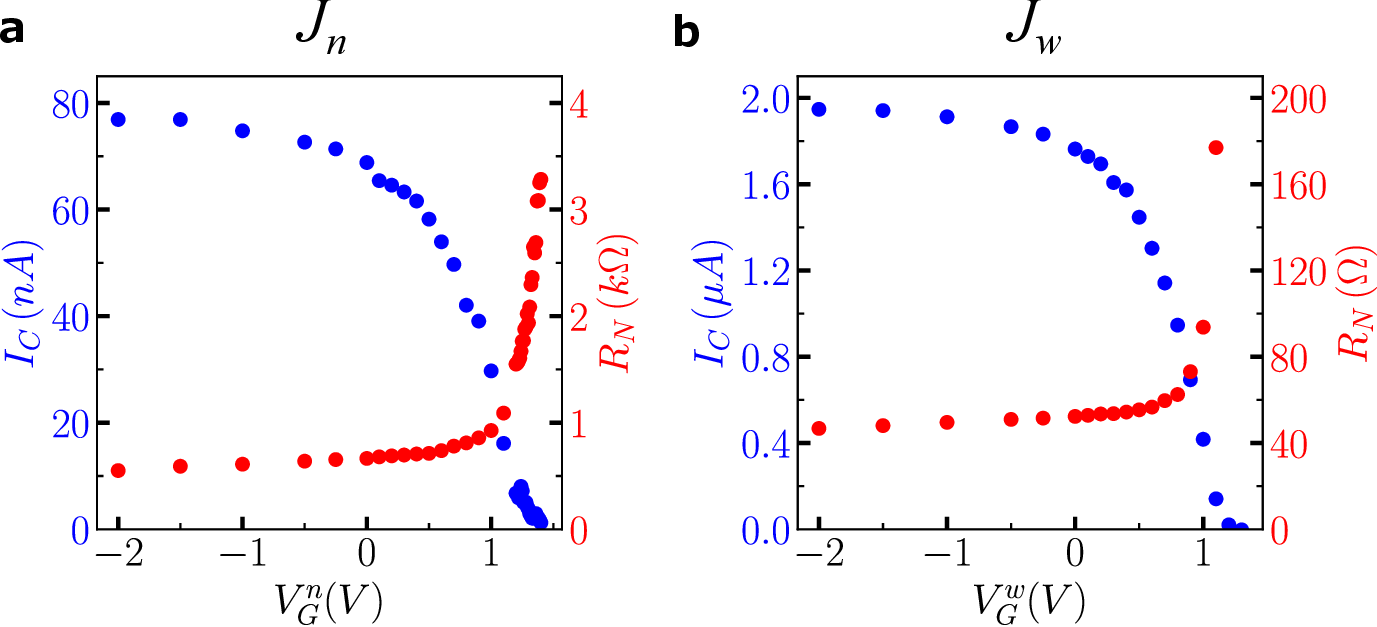}
\caption{\textbf{Critical current and normal resistance of $\mathrm{J^n}$ and $\mathrm{J^w}$.} \textbf{a} The data for the narrow junction $\mathrm{J^n}$ (W=\SI{1}{\um} and L=\SI{300}{\nm}). \textbf{b} The data for the wide junction $\mathrm{J^w}$ (W=\SI{8}{\um} and L=\SI{300}{\nm}).}
\label{Supp_IcRn}
\end{figure}


\section{Model including loop inductance}\label{sup_ind}
We describe a way of computing the critical current of a SQUID made up of JJs with a multi-harmonics CPR and taking into account the inductances of each arms of the loop. The method is adapted from \cite{lecocq_dynamique_2011, dumur_v-shape_2015}.

For simplicity, the CPR of a single junction is defined as follows \cite{beenakker_josephson_1991}:
\begin{equation}
    I(\varphi) = I_C \frac{\overline\tau \sin(\varphi)}{\sqrt{1-\overline\tau \sin^2(\varphi/2)}}
    \label{eq:beenakker_average}
\end{equation}
where $I_C$ is the critical current, $\overline\tau$ is the junction main transparency and $\varphi$ its phase. Although not strictly correct, such description allows to capture the multiple harmonics dependence of the CPR with limited parameters. The corresponding Josephson energy is then written: 
\begin{equation}
    E_J(\varphi) = I_C \frac{2\hbar}{e} \sqrt{1-\overline\tau sin^2(\varphi/2)}
\end{equation}

In the SQUID, the two junctions phase differences respect the condition:
\begin{equation}
    \varphi_\mathrm n-\varphi_\mathrm w = 2\pi \frac{\Phi}{\Phi_0}
\end{equation}
where $\varphi_\mathrm w$ and $\varphi_\mathrm n$ are the phase differences across the two junctions, $\Phi_0$ the flux quantum, $\Phi$ is the flux through the SQUID loop. 

\begin{figure}[H]
\centering
\includegraphics[width=0.35\textwidth]{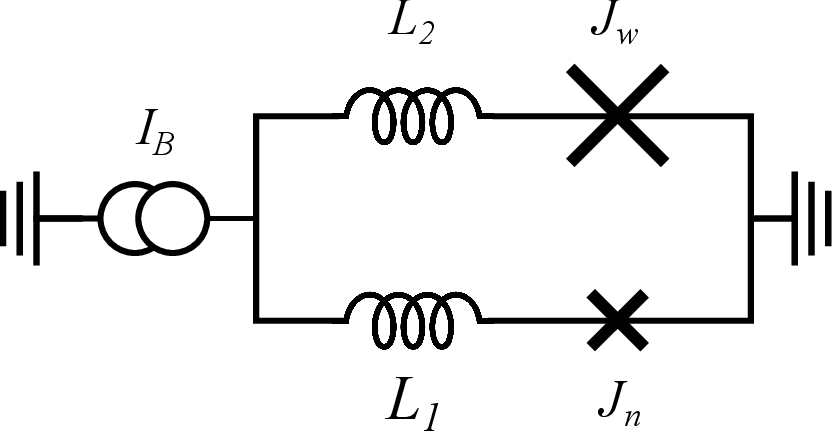}
\caption{\textbf{Equivalent circuit of the SQUID including arm inductances.} $J_\mathrm n$ and $J_\mathrm w$ are the narrow and the wide junctions, $L_\mathrm 1$ and $L_\mathrm 2$ are the inductances of the two arms of the SQUID. The SQUID is current biased by $I_B$.}
\label{electric_draw}
\end{figure}

Furthermore, the two arms loop inductances $L_\mathrm 1$ and $L_\mathrm 2$ (as described in Fig. \ref{electric_draw}) induce a flux screening of the applied external flux $\Phi_\mathrm{ext}$ so that:
\begin{equation}
    \Phi = \Phi_\mathrm{ext} + I_\mathrm{screen}(L_\mathrm 1+L_\mathrm 2)
\end{equation}
where $I_\mathrm{screen}$ is the screening current flowing in the SQUID loop:

\begin{equation}
    I_\mathrm{screen} = \frac{1}{L_\mathrm 1+L_\mathrm 2}\left[\frac{\Phi_0}{2\pi}(\varphi_\mathrm n - \varphi_ \mathrm w)-\Phi_\mathrm{ext}\right]
\end{equation}
Thus, the inductance energy stored in each arm is:
\begin{equation}
    E_{L_\mathrm 1} = \frac{1}{2} L_\mathrm 1 \left(\frac{I_\mathrm B}{2} + I_\mathrm{screen}\right)^2
    \quad \quad \quad
    E_{L_\mathrm 2} = \frac{1}{2} L_\mathrm 2 \left(\frac{I_\mathrm B}{2} - I_\mathrm{screen}\right)^2
\end{equation}
where $I_B$ is the applied bias current. Also, the junctions driven energy are written as follows:
\begin{equation}
    E_B^\mathrm w = -I_\mathrm B \left(\frac{\Phi_0}{2\pi}\right)\varphi_\mathrm w
    \quad \quad \quad
    E_B^\mathrm s = -I_\mathrm B \left(\frac{\Phi_0}{2\pi}\right)\varphi_\mathrm n
\end{equation}
and the total SQUID potential energy can be written as the sum of all the previously derived contributions:
\begin{equation}
    U = -E_J^\mathrm w - E_J^\mathrm n - E_B^\mathrm w - E_B^\mathrm n - E_{L_\mathrm 1} - E_{L_\mathrm 2}
\end{equation}

We define the new phase coordinates x and y respectively for the symmetric and anti-symmetric modes:
\begin{equation}
    x = \frac{\varphi_\mathrm w + \varphi_\mathrm n}{2}
    \quad \quad \quad
    y = \frac{\varphi_\mathrm w - \varphi_\mathrm n}{2}
\end{equation}
and introduce the circuit parameters and the normalized bias current:
\begin{equation}
    \alpha = \frac{I_C^\mathrm w - I_C^\mathrm n}{I_C^\mathrm w + I_C^\mathrm n}
    \quad \quad \quad
    \eta = \frac{L_\mathrm 2 - L_\mathrm 1}{L_\mathrm 1 + L_\mathrm 2}
    \quad \quad \quad
    \beta = \frac{L_\mathrm 1 + L_\mathrm 2}{\Phi_0 / 2\pi (I_C^\mathrm w+I_C^\mathrm n)}
    \quad \quad \quad
    s = \frac{2I_B}{I_C^\mathrm w + I_C^\mathrm s}
\end{equation}

Finally, assuming identical mean transparency $\overline \tau$ in each junction, we get the full expression for the potential:

\begin{eqnarray*}
    U(x,y) & = & U_0 \left[ s(x+\eta y)\Phi_0^2 - \frac{1}{\beta}(\Phi_0 y - \pi \Phi_\mathrm{ext})^2  + \sqrt{2}\Phi_0^2 \left( (1+\alpha)\sqrt{2-\overline\tau+\overline\tau \cos (x-y)} \right.  \right. \\
    & &         \left. \left. + (1-\alpha) \sqrt{2-\overline\tau+\overline\tau \cos (x+y)}  \right) \right]
\end{eqnarray*}

For a given $\Phi_\mathrm{ext}$ and $I_B$, the potential U consist in wells separated by saddle points. When $I_B$ reaches the critical current, these potential wells does not represent stable position any longer. This situation can be described by the following conditions:

\begin{equation}
    \begin{cases}
        \partial_x U(x,y,s) & = 0 \\
        \partial_y U(x,y,s) & = 0 \\
        \partial_{xx} U(x,y,s) \partial_{yy}U(x,y,s) - \partial_{xy}U(x,y,s) \partial_{yx}U(x,y,s) & =0    
    \end{cases}
\label{syst_simu}
\end{equation}

In what follows, the critical current as a function of the external flux is calculated by numerically solving \ref{syst_simu}.

The inductances $L_\mathrm 1$ and $L_\mathrm 2$ have a geometric and a kinetic contribution $L_\mathrm{geo}$ and $L_\mathrm{kin}$. The kinetic inductance per square $L_\mathrm{kin}^\mathrm{S}$ is estimated from the aluminum normal state sheet resistance $R_\mathrm{S} = \SI{1}{\ohm}$ and its superconducting critical temperature $T_\mathrm{C}=\SI{1.54}{\kelvin}$ \cite{mattis_theory_1958,tinkham_introduction_2015}:
\begin{equation}
    L_\mathrm{kin}^S \approx \frac{\hbar}{\pi} \frac{R_S}{1.76 k_B T_C}
\end{equation}
where $k_B$ is the Boltzmann constant. We find $L_\mathrm{kin}^S = \SI{1}{\pico\henry\per\sq}$. To estimate the total loop inductance  $L = L_\mathrm{kin} + L_\mathrm{geo}$, we use a finite elements simulation performed in Sonnet and find $L = \SI{102}{\pico\henry}$. The contribution of the SQUID arm containing the small (resp. wide) junction is $L_\mathrm{1}=\SI{50}{\pico\henry}$ (resp. $L_\mathrm{2}=\SI{52}{\pico\henry}$). Thus the characteristic parameters of the SQUID are:

\begin{equation}
    \beta = \frac{L}{\Phi_0 / 2\pi I_C} = 0.64
    \quad \quad \quad
    \eta = \frac{L_\mathrm 2-L_\mathrm 1}{L_\mathrm 1+L_\mathrm 2} = 0.02
\end{equation}

The SQUID loop inductance and the presence of higher harmonics in the junction CPR may both lead to skewness in the $I_\mathrm{C}$ versus $\Phi$ data. To discriminate their respective contributions in our measurements we use the model described above where booth contributions are taken into account. In all the following fits, the wide junction critical current is set to $I_\mathrm{C}^\mathrm{w}=\SI{2.39}{\uA}$ according to Fig.~\ref{Supp_stats}d and the data used correspond to the $V_\mathrm{G}^\mathrm{n}=V_\mathrm{G}^\mathrm{w}=\SI{-2}{\V}$ configuration.

\begin{figure}[H]
\centering
\includegraphics[width=0.85\textwidth]{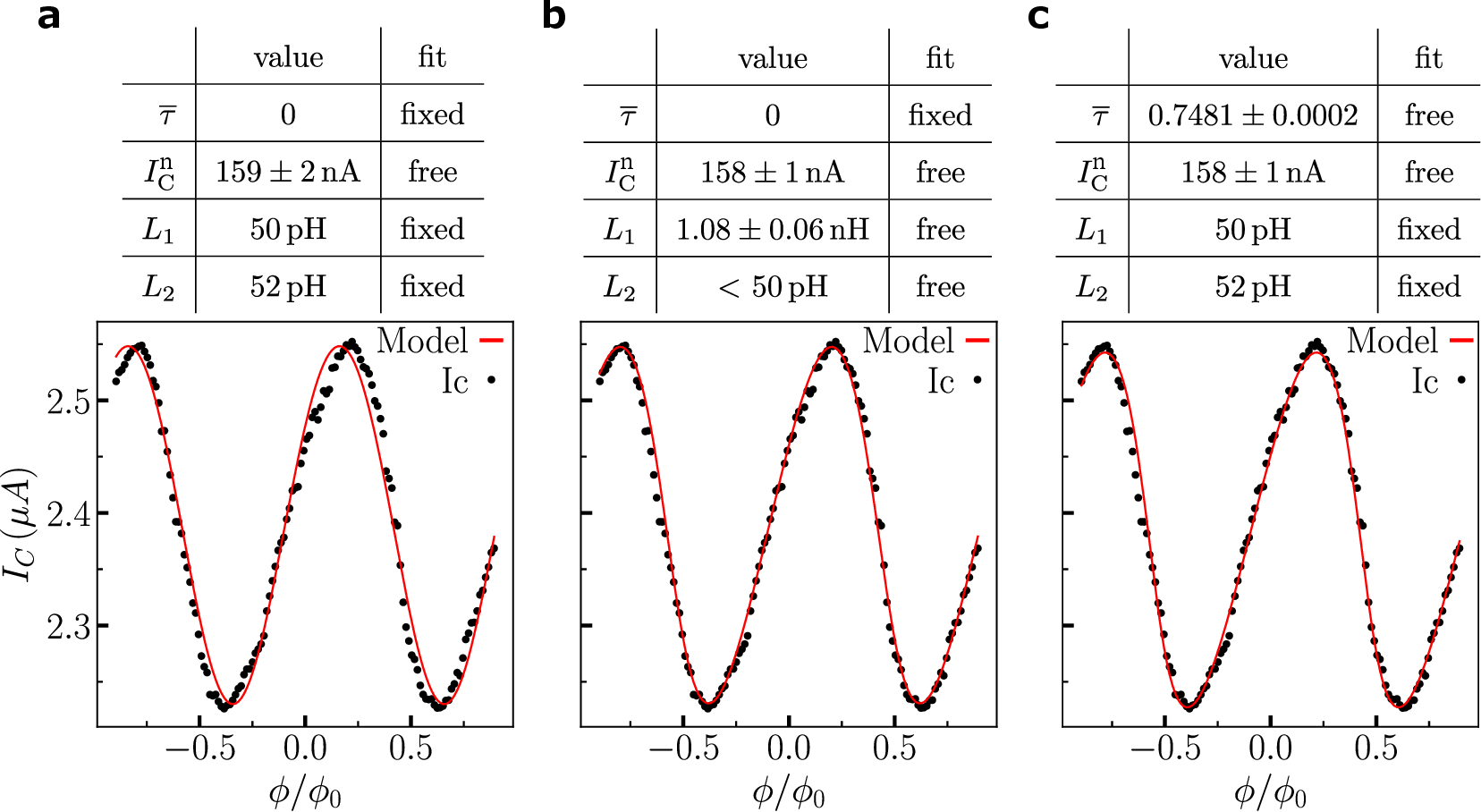}
\caption{\textbf{CPR data fitting taking the SQUID loop inductance into account.} \textbf{a}, The junctions CPR are sinusoidal, the loop inductance is set to the value calculated with Sonnet and $I_C^\mathrm n$ is the only fitting parameter. \textbf{b}, The junctions CPR are sinusoidal, the two arms inductances $L_\mathrm 1$ and $L_\mathrm 2$ and $I_C^\mathrm n$ are the fitting parameters. \textbf{c}, The loop inductance is set to the value calculated with Sonnet, the junctions CPR follows the SNS short junction model, their transparency $\overline\tau$ and $I_C^n$ are the fitting parameters.}
\label{Supp_fit}
\end{figure}

\begin{figure}[H]
\centering
\includegraphics[width=0.71\textwidth]{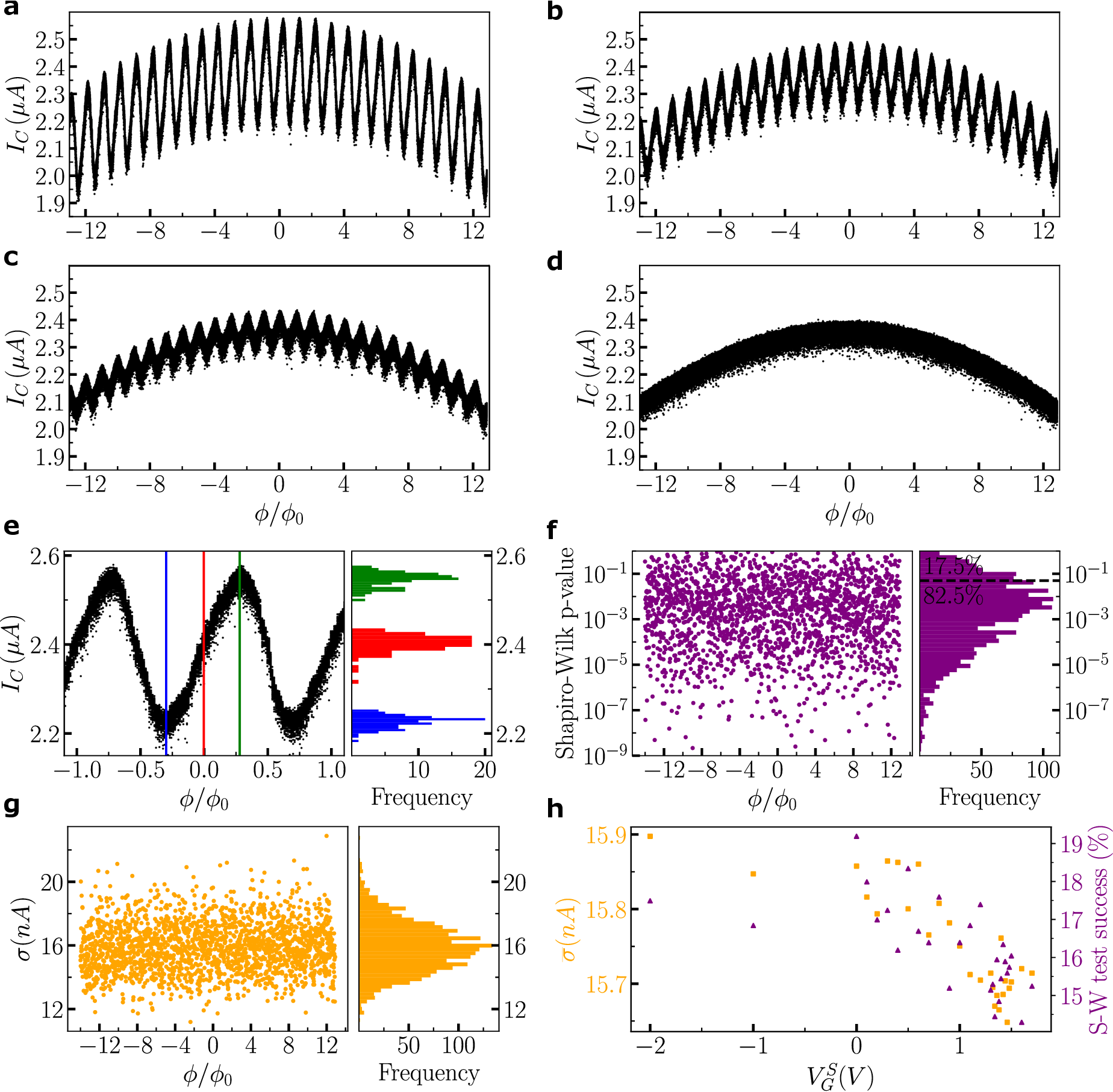}
\caption{\textbf{Asymmetric SQUID raw critical current data and statistical analysis.}}
 \label{Supp_stats}
\end{figure}

First, the transparency is fixed to zero meaning that we consider the sinusoidal CPR regime for the two junctions. The arms inductances are set to the values calculated with Sonnet and $I_\mathrm{C}^\mathrm{n}$ is the only free parameter. The fit, shown in Fig.~\ref{Supp_fit}a, results in a $\chi^2=\SI{3.8e10}{}$ and the CPR skewness can poorly be reproduced with this loop inductance values.

Second, Fig.~\ref{Supp_fit}b shows the same fit except that the inductances are free parameters. The fit results in a $\chi^2=\SI{1.3e10}{}$ and the CPR skewness can be reproduced taking a tremendous and asymmetric loop inductance, $L_\mathrm 1=\SI{1.08}{\nano\henry}$ and $L_\mathrm 2<\SI{50}{\pico\henry}$. However, this inductance is completely at odds with the Sonnet simulations.

Third, Fig.~\ref{Supp_fit}c shows a parameter set reproducing the CPR skewness where the inductances are set to the values calculated before and where the junctions CPR are modeled by the SNS short junction theory (\ref{eq:beenakker_average}). The average transparency $\overline{\tau}$ and $I_\mathrm{C}^\mathrm{n}$ were the only free parameters and $\chi^2=\SI{1.4e10}{}$.  

To conclude, the SQUID loop inductance cannot by itself explain the data and the use of a multi harmonics CPR model is required to reproduce the observed skewness.

\newpage

\section{Critical current statistics}\label{sup_stat}

In Fig. S2a,b,c,d, each measured critical current is represented by a black spot, 100 $I_C$ measurements are performed for each flux value $\Phi$. The wide junction gate voltage is kept at $V_G^\mathrm w=\SI{-2}{\V}$ while the small junction gate voltage is $V_G^\mathrm n=\SI{-2}{\V}$ in \textbf{a}, $V_G^\mathrm n=\SI{1.1}{\V}$ in \textbf{b}, $V_G^\mathrm n=\SI{1.3}{\V}$ in \textbf{c}, 
 $V_G^\mathrm n=\SI{1.5}{\V}$ in \textbf{d}. The oscillations are due to the SQUID effect and associated with the small junction CPR, they disappear near pinch-off (i.e. when $V_G^\mathrm n$ is near $V_{th} \approx \SI{1.5}{\V}$). The background curvature is associated with the Fraunhofer effect over the wide junction and is independent from $V_G^\mathrm n$. Fig. S2e shows a zoom in around $\Phi=0$ at $V_G^\mathrm n=-2V$. The $I_C$ distributions at $\Phi/\Phi_0=-0.3$, $\Phi/\Phi_0=0$ and $\Phi/\Phi_0=0.28$ are shown in the right panel. In Fig. S2f, A Shapiro-Wilk test for normality is performed on the $I_C$ distribution at each flux value for $V_G^\mathrm n=\SI{-2}{\V}$. The resulting p-values are plotted as a function of the applied flux. The normality of the $I_C$ distribution appears to be uncorrelated with the applied flux. The right panel shows the distribution of the Shapiro-Wilk p-values. We notice that for 82.5\% of the $\Phi$ values the p-value is below the usual threshold ($5.10^{-2}$), indicating the non normality of the $I_C$ distribution for these flux values. In Fig. S2g, for $V_G^\mathrm n=\SI{-2}{\V}$, the standard deviation $\sigma$ of the $I_C$ distribution is plotted as a function of the applied flux. The standard deviation appears to be uncorrelated with the applied flux. The right panel shows the $\sigma$ distribution for all $\Phi$ values. In Fig. S2h, the average standard deviation $\overline{\sigma}$ as function of the small junction gate voltage $V_G^\mathrm n$ is plotted in yellow. The $\overline{\sigma}$ does not vary by more than 2\% over the entire gate voltage range meaning that the critical current dispersion amplitude is not correlated with the small junction gate voltage value $V_G^\mathrm n$. At each gate voltage $V_G^\mathrm n$, the Shapiro-Wilk test for normality is performed for each flux and the fraction of p-values found to be above the $5.10^{-2}$ threshold is reported in purple. Again, this quantity does not vary by more than 5\% over the entire $V_G^n$ range, implying that the nature of the $I_C$ distribution is not correlated with the applied $V_G^\mathrm n$. However, the fact that the critical current distribution does not follow a normal law has already been reported \cite{holmes_non-normal_2017}, but in our case is probably due to the measurement technique (using a counter without signal integration). The detection of the normal state resistance transition is sensitive to low current false positives induced by noise above the detection threshold in the measured voltage.

\newpage

\section{Shapiro pattern at different gate voltages and temperatures}\label{sup_shap}

\begin{figure}[H]
\centering
\includegraphics[width=0.84\textwidth]{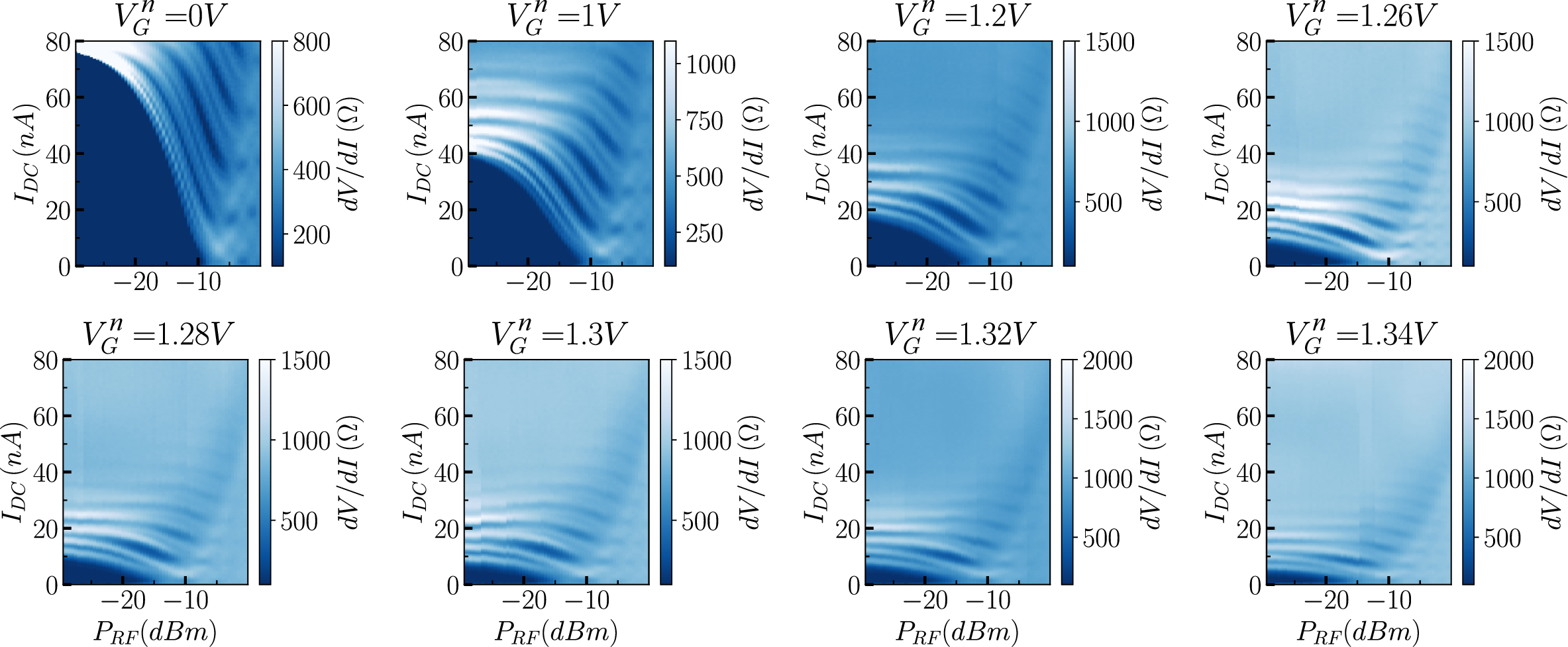}
\caption{\textbf{Shapiro pattern of the asymmetric SQUID small junction at different gate voltages.} The wide junction is kept in depletion ($V_G^\mathrm w=\SI{4}{\V}$) so that the small junction is the only one to be probed. The experimental setup is the same as in part IV with $f=\SI{3.05}{\giga\Hz}$ and here the measurement is reproduced at different $V_G^\mathrm n$.}
\label{Supp_shap_gate}
\end{figure}

The measurement of part IV is reproduced at different $V_G^\mathrm n$ keeping the wide junction in depletion ($V_G^\mathrm w=\SI{4}{\V}$). Fig. \ref{Supp_shap_gate} shows the different Shapiro patterns at various $V_G^\mathrm n$ with $f=\SI{3.05}{\giga\Hz}$. The higher harmonics of the CPR disappear when the gate voltage is close to the threshold voltage ($V_{th} \approx \SI{1.5}{\V}$) (see Sec. III). Thus, one can expect the half integer steps to disappear before the integer ones when $V_G^\mathrm n$ gets close to the threshold. A rigorous quantitative analysis of such a phenomenon is challenging in this system because the Shapiro steps are rounded. In fact, the signatures of the half-integer and integer steps in the $dV/dI$ measurement often overlap and mix, especially near threshold. 

\begin{figure}[H]
\centering
\includegraphics[width=0.63\textwidth]{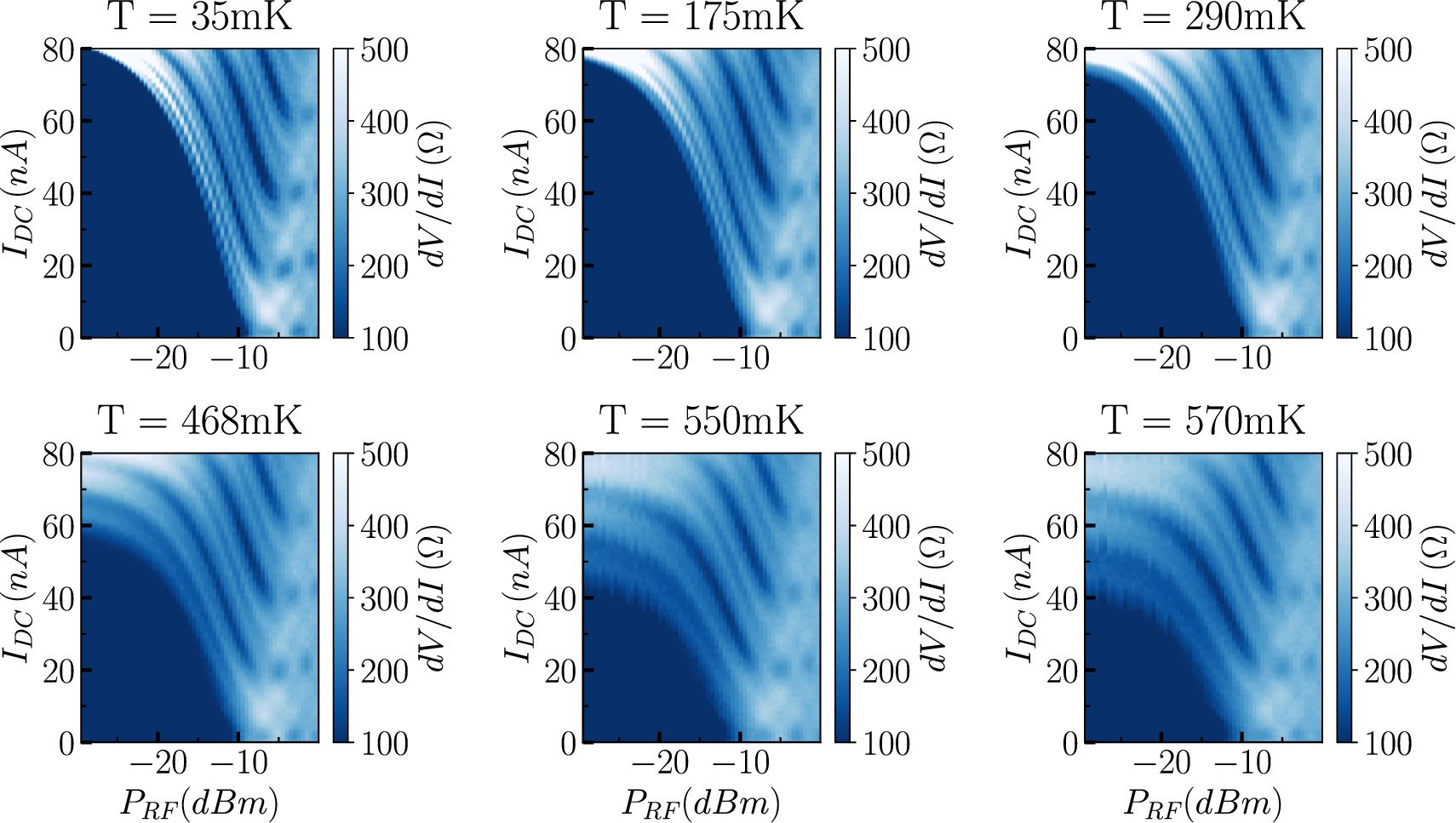}
\caption{\textbf{Shapiro pattern of the asymmetric SQUID small junction at various temperatures.} The wide junction is kept in depletion ($V_G^\mathrm w=\SI{4}{\V}$) so that the small junction (kept in accumulation $V_G^\mathrm n=\SI{-2}{\V}$) is the only one to be probed. The experimental setup is the same as in part IV with $f=\SI{3.05}{\giga\Hz}$ and the measurement is reproduced at different temperatures.}
\label{Supp_shap_temp}
\end{figure}

Then, we investigate the temperature dependence of the small junction Shapiro pattern in the accumulation regime ($V_G^\mathrm n = \SI{-2}{\V}$ and $V_G^\mathrm w=\SI{4}{\V})$. Fig. \ref{Supp_shap_temp} shows the different Shapiro patterns at various temperatures with $f=\SI{3.05}{\giga\Hz}$. As the temperature increases, the Shapiro steps signatures in the $dV/dI$ are broadened and blurred. There is no clear enhancement of the half-integer steps with increasing temperature, which could be associated with an out-of-equilibrium effect \cite{dubos_coherent_2001}.

\newpage

\section{SQUID effect oscillations in the symmetric device}\label{sup_squid}

\begin{figure}[H]
\centering
\includegraphics[width=0.95\textwidth]{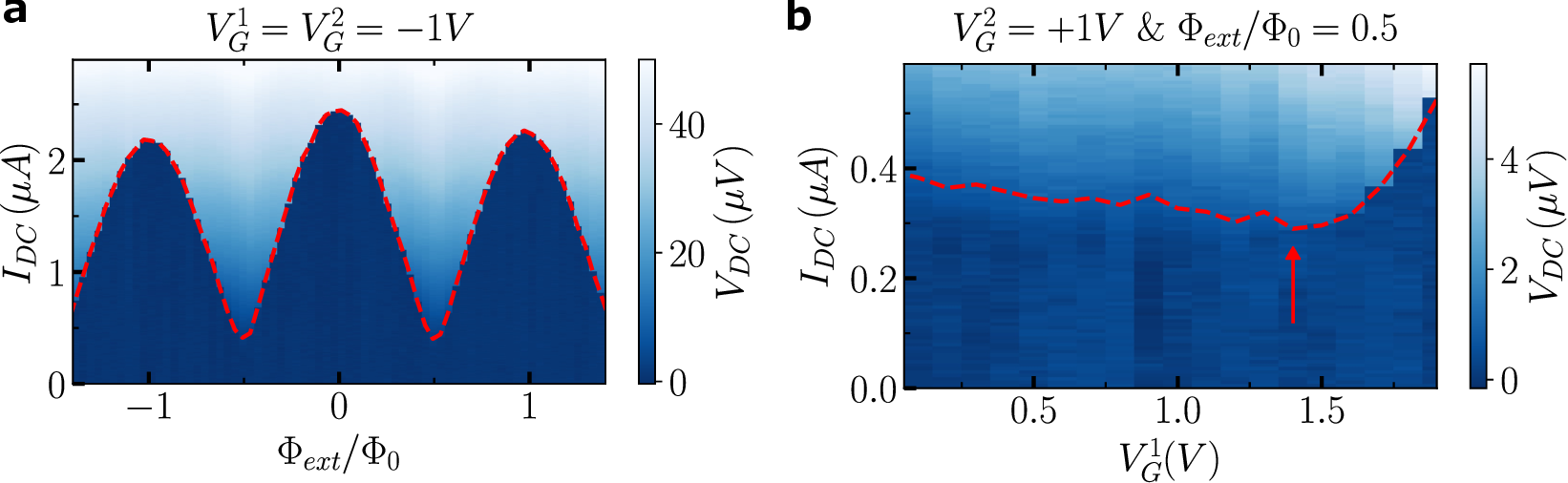}
\caption{\textbf{Symmetric SQUID oscillations and I\textsubscript{C} minimization at half the flux quantum.} \textbf{a}, Measured voltage $V_{DC}$ as a function of the bias current $I_{DC}$ and the applied magnetic flux $\Phi_{ext}$. The two gate voltages are kept at \SI{-1}{\V}. The red dotted line follows the critical current $I_C$ defined by a voltage threshold $V_{th}=\SI{1.5}{\uV}$. \textbf{b}, Measured voltage as a function of the bias current and the $J_\mathrm 1$ gate voltage $V_G^\mathrm 1$ while the applied magnetic flux is held at half flux quantum and $V_G^\mathrm 2=\SI{1}{\V}$. The red dotted line follows $I_C$ defined as before and the red arrow points its minimum where the balanced SQUID regime is reached.}
\label{Supp_squid_effect}
\end{figure}

The usual SQUID effect oscillations are observed in the symmetric SQUID device where $J_\mathrm 1$ and $J_\mathrm 2$ are W=\SI{1}{\um} wide and L=\SI{300}{\nm} long. The SQUID effect experiment is performed with the two junctions in accumulation ($V_G^\mathrm 1=V_G^\mathrm 2=\SI{-1}{\V}$). In Fig. \ref{Supp_squid_effect}a, we show the measured voltage as a function of the bias current and the applied magnetic flux $\Phi_{ext}$. The red dotted line follows the critical current $I_C$ defined by a threshold voltage $V_{th}=\SI{1.5}{\uV}$. The periodicity of the $I_C$ maxima and minima is used to calibrate the relationship between the applied perpendicular magnetic field and the magnetic flux through the SQUID. Furthermore, the more balanced the SQUID is (i.e. the critical currents of the two junctions are equal), the lower $I_C$ will be at half the flux quantum ($\Phi_{ext}/\Phi_0=0.5$). Thus, one way to find the gate voltage combinations that allow reaching the symmetric regime is to try to minimize $I_C$ at $\Phi_{ext}/\Phi_0=0.5$. In Fig. \ref{Supp_squid_effect}b, we show the measured voltage as a function of the bias current and the $J_\mathrm 1$ gate voltage $V_G^\mathrm 1$. The applied flux is held at half the flux quantum and $V_G^\mathrm 2=\SI{1}{\V}$. The red dotted line follows $I_C$ defined as before. The minimum $I_C$ is reached at $V_G^\mathrm 1 \approx \SI{1.4}{\V}$ which means that the SQUID is operated in a symmetric regime for the gate combination $V_G^1=\SI{1.4}{\V}$ \& $V_G^2=\SI{1}{\V}$. This result is consistent with the Fig. 4d where we see that this gate combination is one of the possibilities to reach the balanced SQUID regime.

\section{Critical current in the balanced SQUID regime}\label{sup_sweet_line}

\begin{figure}[H]
\centering
\includegraphics[width=0.5\textwidth]{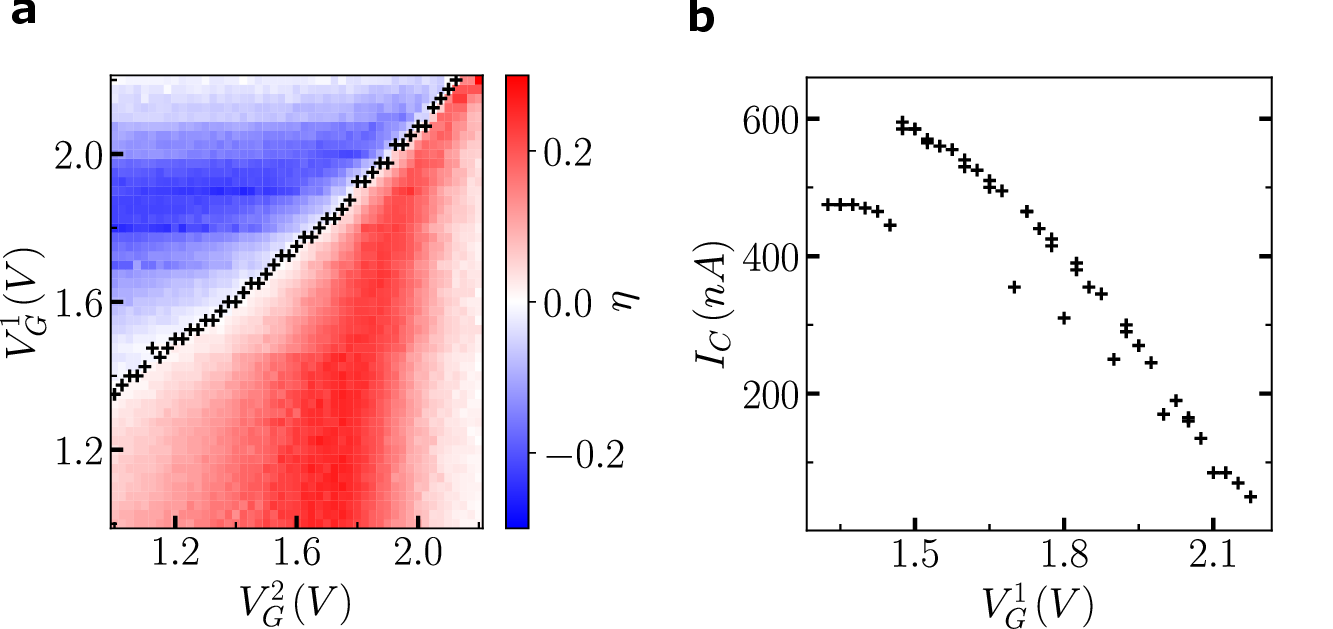}
\caption{\textbf{Balanced SQUID line.} \textbf{a}, Diode efficiency $\eta$ at fixed applied flux $\Phi_{ext}/\Phi_0 = 0.45$ as a function of the two gate voltages $V_G^\mathrm 1$ and $V_G^\mathrm 2$ of the symmetric SQUID. The $\eta$ minima are marked by the black crosses, they define to the sweet line. \textbf{b}, SQUID critical current $I_C$ along the line as a function of the gate voltages $V_G^\mathrm 1$ while $V_G^\mathrm 2$ is adjusted to be in the zero diode efficiency regime.}
\label{Supp_sweet_line}
\end{figure}

In Fig. \ref{Supp_sweet_line}a we show the diode efficiency $\eta$ at fixed applied flux $\Phi_{ext}/\Phi_0 = 0.45$ (same as in Fig. 4d). When the SQUID is balanced, the diode efficiency goes to zero and so the line is defined by the diode efficiency minima and marked by the black crosses. The SQUID critical current $I_C$ along this line is plotted on Fig. \ref{Supp_sweet_line}b as a function of the two gate voltages. $I_C$ decreases as the gate voltages approach pinch-off. The $I_C$ abrupt fluctuations are not related to gate voltage fluctuations since there is not such pattern in the sweet line versus gates (Fig. \ref{Supp_sweet_line}a). Magnetic flux fluctuations smaller than $\Phi_0/20$ could explain the phenomenon.

\section{Influence of the diode effect on the Shapiro patterns}\label{sup_diode_shap}

\begin{figure}[H]
\centering
\includegraphics[width=0.48\textwidth]{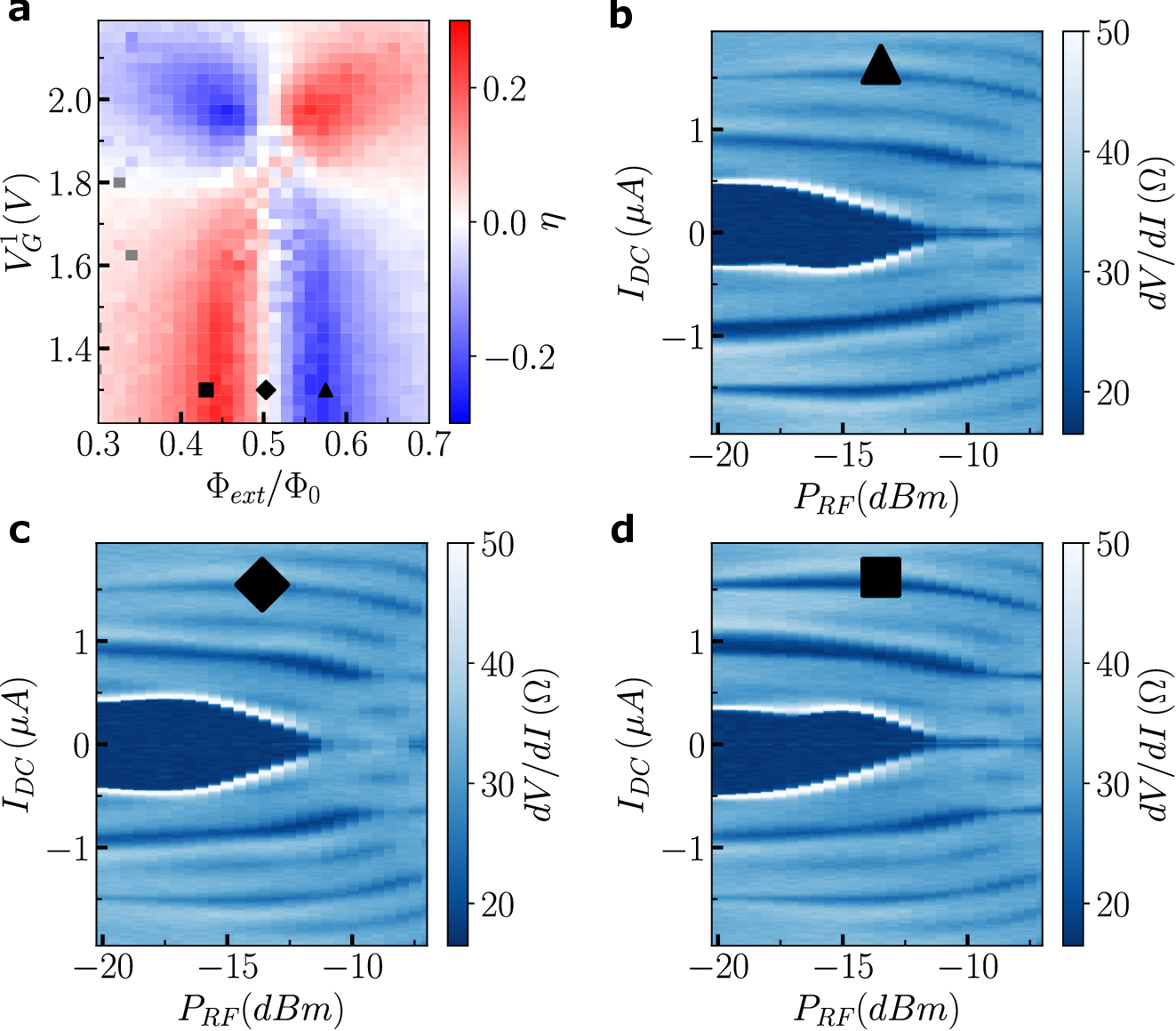}
\caption{\textbf{Shapiro patterns at negative, null and positive diode efficiency.} \textbf{a}, Zoom in the diode efficiency pattern shown in Fig. 4b where $\eta$ is plotted as function of $J_\mathrm 1$ gate voltage ($V_G^\mathrm 1$) and the applied magnetic flux $\Phi_{ext}$. $V_G^\mathrm 2$ is kept at \SI{1.75}{\V}. The three signs ($\blacksquare, \blacklozenge, \blacktriangle$) indicate the gate and flux configurations at which the Shapiro patterns are measured. \textbf{b},\textbf{c},\textbf{d}, Shapiro patterns measured by current biasing always from zero to positive/negative values at $V_G^\mathrm 1=\SI{1.3}{\V}$ and $V_G^\mathrm 2=\SI{1.75}{\V}$. The flux biasing is kept at $\Phi_{ext}/\Phi_0=0.57$ in \textbf{b}, $\Phi_{ext}/\Phi_0=0.5$ in \textbf{c} and $\Phi_{ext}/\Phi_0=0.43$ in \textbf{d}.}
\label{Supp_shap_diode}
\end{figure}

In Fig. \ref{Supp_shap_diode}, we study the Shapiro pattern in the unbalanced regime ($V_G^\mathrm 1=\SI{1.3}{\V}$ \& $V_G^\mathrm 2=\SI{1.75}{\V}$) for three different applied fluxes corresponding to a negative diode efficiency (Fig. \ref{Supp_shap_diode}b), a zero diode efficiency (Fig. \ref{Supp_shap_diode}c) and a positive diode efficiency (Fig. \ref{Supp_shap_diode}d). The bias current ramps always start from zero toward positive/negative values so that the forward/backward current configurations are probed. We note that the Shapiro steps width are also affected by the diode effect. The kink in the backward (resp. forward) $I_C$ power dependency observed in Fig. \ref{Supp_shap_diode}b (resp. Fig. \ref{Supp_shap_diode}d) remains unexplained.

\section{Additional Shapiro measurements around the second harmonic enhancement sweet spot}\label{sup_around_sweet_spot}

During a new cool down of the same device, we focus on the center of the diode efficiency pattern shown in Fig. 4b and notice that it is not exactly symmetric. Thus, we investigate the Shapiro patterns in this region and find that the presence or absence of integer Shapiro steps correlates mainly with the applied magnetic flux and not with the gate voltage fine tuning.

\begin{figure}[H]
\centering
\includegraphics[width=0.70\textwidth]{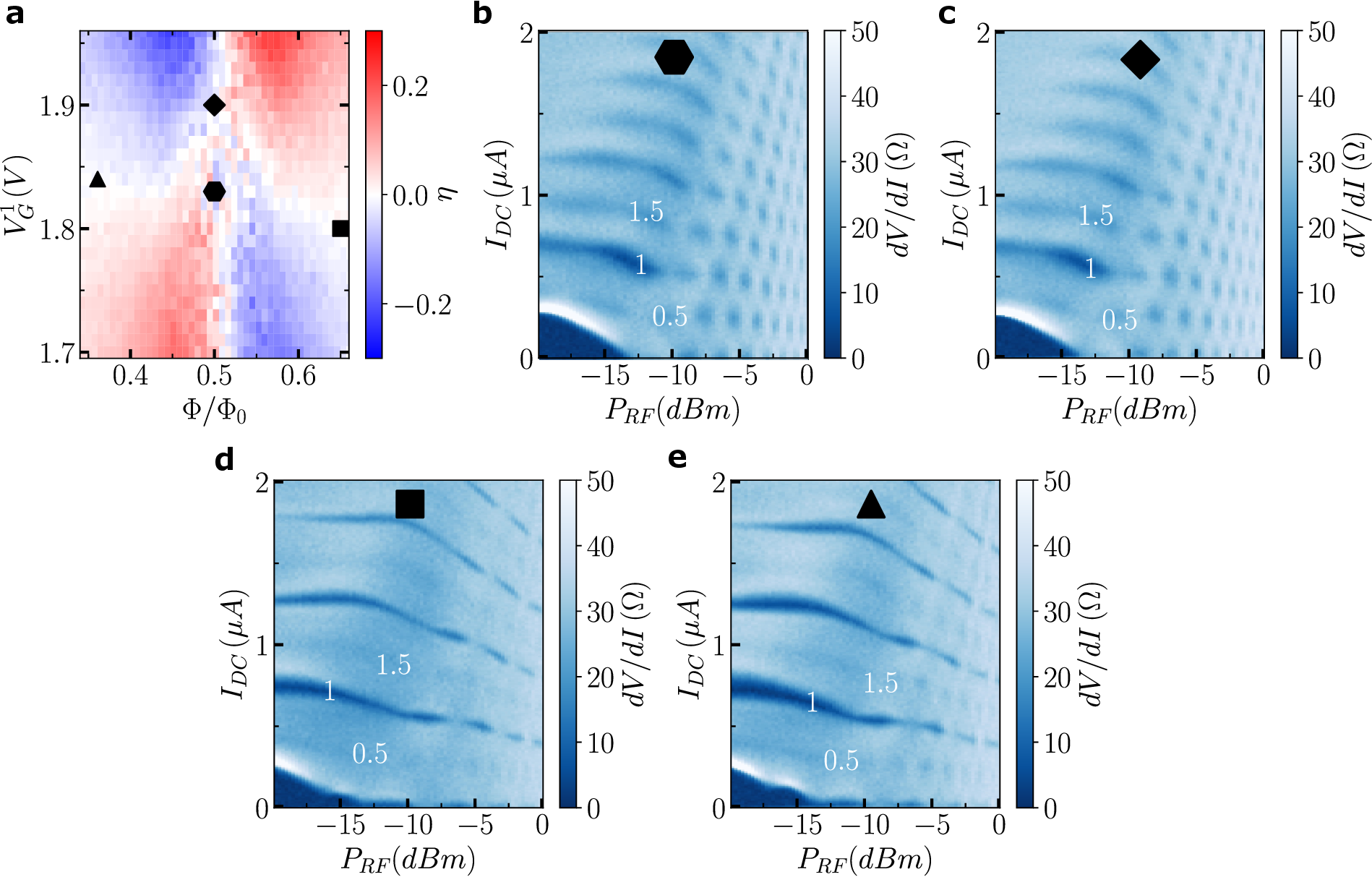}
\caption{\textbf{Shapiro patterns around the balanced SQUID and half flux sweet spot configuration.} \textbf{a}, Diode efficiency $\eta$ as a function of the $J_\mathrm 1$ gate voltage $V_G^\mathrm 1$ and the applied magnetic flux $\Phi_{ext}$. $V_G^\mathrm 2$ is kept at \SI{1.75}{\V}. The four signs indicate the gate and flux configurations at which the Shapiro patterns are measured.\textbf{b},\textbf{c},\textbf{d},\textbf{e}, All the Shapiro patterns are measured at $V_G^\mathrm 2=\SI{1.75}{\V}$. $J_\mathrm 1$ gate voltage and magnetic flux bias are $V_G^\mathrm 1=\SI{1.83}{\V}$ and $\Phi_{ext}/\Phi_0=0.5$ in \textbf{b}, $V_G^\mathrm 1=\SI{1.9}{\V}$ and $\Phi_{ext}/\Phi_0=0.5$ in \textbf{c}, $V_G^\mathrm 1=\SI{1.8}{\V}$ and $\Phi_{ext}/\Phi_0=0.65$ in \textbf{d}, $V_G^\mathrm 1=\SI{1.84}{\V}$ and $\Phi_{ext}/\Phi_0=0.36$ in \textbf{e}. }
\label{Supp_shap_sweet_spot_tuning}
\end{figure}

\newpage

\section{Symmetric SQUID current phase relation}\label{sup_simu}

\begin{figure}[H]
\centering
\includegraphics[width=0.80\textwidth]{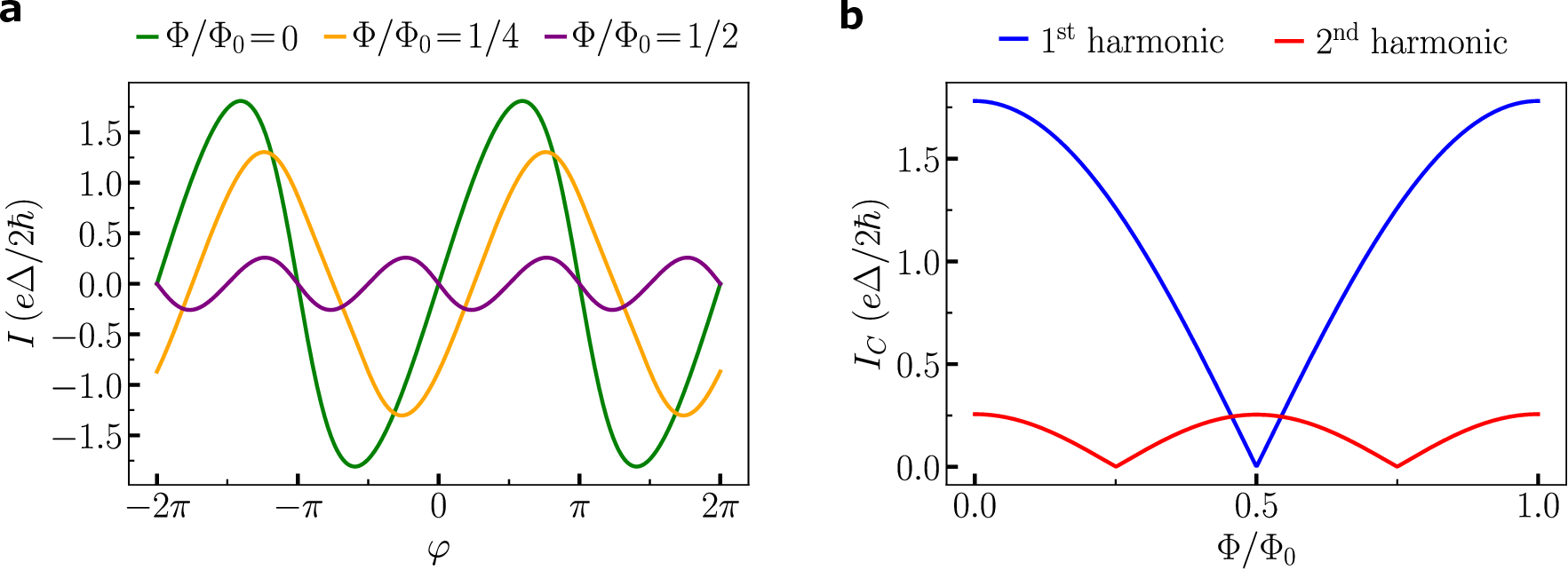}
\caption{\textbf{Computation of a symmetric SQUID CPR for a single channel short junction model with a transparency $\mathbf{\tau=0.7}$.} \textbf{a}, The SQUID CPR for different flux bias values. The normalized current $I$ flowing through the SQUID is plotted as a function of the phase of one of the two junctions. \textbf{b}, The two first harmonic contributions to the SQUID critical current $I_C$ of the same system plotted as a function of the normalized magnetic flux $\Phi/\Phi_0$.}
\label{Supp_squid_simu}
\end{figure}

We compute the current phase relation (CPR) of a symmetric SQUID. The two junction CPRs are modeled by the short junction single channel at zero temperature formula \cite{haberkorn_theoretical_1978}, which gives the current $I$ as a function of the junction phase $\varphi$ for a given transparency $\tau$:

\begin{equation}
    I(\varphi) = \frac{e \Delta}{2\hbar} \frac{\tau sin(\varphi)}{\sqrt{1-\tau sin^2(\varphi/2)}}
\end{equation}
where $e$ is the elementary charge, $\hbar$ the reduced Plank constant and $\Delta$ the superconducting gap.  Thus, the total SQUID CPR is the sum of the two junction CPRs considering the fluxoid relation:

\begin{equation}
    I_{SQUID}(\varphi) = I_\mathrm 1(\varphi) + I_\mathrm 2(\varphi-2\pi\Phi/\Phi_0)
\end{equation}
where $I_\mathrm 1$ (resp. $I_\mathrm 2$) is the CPR of the junction $J_\mathrm 1$ (resp. $J_\mathrm 2$), $\varphi$ is the $J_\mathrm 1$ phase, $\Phi$ is the applied magnetic flux through the SQUID and $\Phi_0=h/2e$ is the flux quantum. Fig. \ref{Supp_squid_simu} shows the computed CPR and its harmonic decomposition for a SQUID where the junctions transparency are $\tau=0.7$. 

At integer flux quantum, the SQUID CPR contains odd and even harmonics but at half flux quantum, the odd harmonics vanish and so the fundamental frequency doubles. Furthermore, at quarter flux quantum, we notice that the even harmonics vanish. Considering only the first two harmonics (the amplitudes of the next ones are usually negligible), the SQUID behaves as a so-called $sin(2\varphi)$ Josephson element at half the flux quantum, as a pure $sin(\varphi)$ at quarter flux quantum and as a multi harmonics element at integer flux quantum. 

\end{document}